\documentclass[conference,compsoc]{template/IEEEtran}
\IEEEoverridecommandlockouts
\usepackage[T1]{fontenc}
\usepackage{microtype}
\usepackage[sort]{cite}
\usepackage[hyphens]{url}
\usepackage{graphicx}
\usepackage[dvipsnames]{xcolor}
\usepackage{balance}
\usepackage{stfloats}
\usepackage[hidelinks]{hyperref}
\usepackage[capitalize,nameinlink,nosort]{cleveref}
\usepackage{xspace}
\usepackage[per-mode=symbol,mode=text,detect-all]{siunitx}
\DeclareSIUnit{\million}{M\@}
\DeclareSIUnit{\thousand}{k\@}
\usepackage[inline]{enumitem}
\usepackage{booktabs}
\usepackage{tabularx}
\usepackage{makecell}
\usepackage{multirow}
\usepackage{array}

\newcommand{\tabitem}{\textbullet~}

\usepackage{circledsteps}
\pgfkeys{/csteps/inner ysep=3pt}
\pgfkeys{/csteps/inner xsep=3pt}
\usepackage{pifont}
\newcommand{\cmark}{\textcolor{ForestGreen}{\ding{51}}}
\newcommand{\xmark}{\textcolor{Red}{\ding{55}}}
\newcommand{\manualmark}{\textcolor{BurntOrange}{\ding{45}}}
\newcommand{\crashmark}{\textcolor{Red}{\ding{61}}}

\newcommand{\impH}{\textcolor{BrickRed}{\textbf{H}}}
\newcommand{\impM}{\textcolor{YellowOrange}{\textbf{M}}}
\newcommand{\impL}{\textcolor{OliveGreen}{\textbf{L}}}

\usepackage{listings}
\crefname{lstlisting}{Listing}{Listings}
\Crefname{lstlisting}{Listing}{Listings}

\definecolor{mygreen}{rgb}{0,0.6,0}
\definecolor{myblue}{rgb}{0.1,0.1,0.8}
\definecolor{mygray}{rgb}{0.498,0.498,0.498}
\definecolor{backcolour}{rgb}{0.95,0.95,0.92}

\newcommand{\nostep}[1]{\Circled{#1}}
\newcommand{\step}[1]{Category~\nostep{#1}}
\newcommand{\steptwo}[1]{Category~\nostep{#1}}

\newcommand{\arxiv}{arXiv\xspace{}}
\newcommand{\arXiv}{\arxiv{}}
\newcommand{\alcpp}{ALC-NG\xspace{}}

\newcommand{\ie}{i.e.,\xspace}
\newcommand{\eg}{e.g.,\xspace}
\newcommand{\Eg}{E.g.,\xspace}
\newcommand{\cf}{cf.\@}

\def\BibTeX{{\rm B\kern-.05em{\sc i\kern-.025em b}\kern-.08em
	T\kern-.1667em\lower.7ex\hbox{E}\kern-.125emX}}

\begin{document}

\bstctlcite{IEEEexample:BSTcontrol}

\title{Hidden Secrets in the \arxiv{}: Discovering, Analyzing, and Preventing\\ Unintentional Information Disclosure in Source Files of Scientific Preprints} %

\author{
	\IEEEauthorblockN{%
		Jan Pennekamp\IEEEauthorrefmark{1}\textsuperscript{,\textsection},
		Johannes Lohmöller\IEEEauthorrefmark{1}\textsuperscript{,\textsection},
		David Schütte\IEEEauthorrefmark{1},
		Joscha Loos\IEEEauthorrefmark{1},
		Martin Henze\IEEEauthorrefmark{3}\textsuperscript{,}\IEEEauthorrefmark{5}
	}
	\IEEEauthorblockA{\IEEEauthorrefmark{1}\textit{Communication and Distributed Systems}, RWTH Aachen University, Germany $\cdot$
	\{lastname\}@comsys.rwth-aachen.de}
	\IEEEauthorblockA{\IEEEauthorrefmark{3}\textit{Security and Privacy in Industrial Cooperation}, RWTH Aachen University, Germany $\cdot$
	\{lastname\}@spice.rwth-aachen.de}
	}

\maketitle
\begingroup
\renewcommand\thefootnote{\textsection}
\footnotetext{Equal contribution.\hspace{2em}$\mathparagraph$. Also with Fraunhofer FKIE.}
\endgroup

\begin{abstract}
Preprints are essential for the timely and open dissemination of research.
\arxiv{}, the most widely used preprint service, takes the idea of open science one step further by not only publishing the actual preprints but also \LaTeX{} sources and other files used to create them.
As known from other contexts, such as GitHub repositories, and anecdotally exemplified for \arxiv{}, making source code publicly available risks disclosing otherwise ``hidden'' information.
Consequently, the public availability of paper sources raises the question of how much sensitive content is (unintentionally) disclosed through them.

In this paper, we systematically answer this question for all \qty{2.7}{\million} \arxiv{} submissions with available source files across three dimensions of source file-induced information disclosure:
\begin{enumerate*}[label=\nostep{\arabic*}{}]
	\item inclusion of unnecessary files,
	\item metadata embedded in files, and
	\item irrelevant content in files such as source code comments.
\end{enumerate*}
Our analysis reveals that nearly every \arXiv{} submission contains some form of ``hidden'' information.
Notable findings range from links to editable web documents for internal coordination over API and private keys to complete Git histories.

While different tools promise to remove such information from source files, we show that they fail to reliably achieve the intended cleaning functionality. 
To mitigate this situation, we provide \alcpp{} to comprehensively remove files, metadata, and comments that are not needed to compile a \LaTeX{} paper.
\end{abstract}


\section{Introduction}
\label{sec:introduction}

Preprints, \ie{} early versions of academic papers already made available before formal peer review~\cite{McKiernan2000arXiv.org:}, are an indispensable tool for the timely and open dissemination of academic research, with first research fields such as machine learning even shifting to a preprint-first ``publication'' model~\cite{citeordie2015Lets}.
Since its inception in 1991~\cite{McKiernan2000arXiv.org:}, \arxiv{}~\cite{Cornell-Tech1991arXiv.org} has established itself as the largest and most frequently used preprint service~\cite{Xieetal2021Is} and is widely accepted in disciplines such as computer science, physics, and mathematics, with \num{>20000} preprints published every month~\cite{arXiv2014Monthly-Submissions}.
In addition to the PDF of the preprint, \arxiv{} also publishes the source code of all papers written using \LaTeX{} ($\approx$\qty{93}{\percent} of submissions).
While initially intended to enable the reuse of notations and provide learning opportunities~\cite{arXiv2005Requiring}, publishing preprint source code enables research such as citation prediction~\cite{Gehrkeetal2003Overview}, studying inconsistencies in documents produced using \LaTeX{}~\cite{Tanetal2024Inconsistencies}, or reconstructing \LaTeX{} sources from PDFs~\cite{Dengetal2019Challenges,Jiangetal2025LATTE:}.

\begin{figure}[t]
	\centering
	\includegraphics[trim=0 0 0 0,clip]{figures/teaser-example}
	\vspace{-1.75em}
	\caption{%
		\LaTeX{} sources might contain \textcolor{red}{comments} or \textcolor{mygray}{dangling files} that are irrelevant to or unused in the compiled PDF, respectively.
		When being distributed regardless, \eg{} via \arxiv{}, they disclose (sensitive) information.
	}
	\label{fig:teaser-example}
\end{figure}

However, besides providing many benefits~\cite{Brinnetal2022A}, making the source code of academic papers publicly available also introduces substantial information disclosure risks~\cite{arXiv2005Requiring}.
As exemplarily highlighted in \cref{fig:teaser-example}, the two main risks stem from
\begin{enumerate*}[label=(\roman*)]
    \item irrelevant content, \eg{} in the comment sections of \LaTeX{} files, and
    \item unused files within the uploaded archive that are not needed to build the paper.
\end{enumerate*}
Notably, these risks come on top of information disclosure through metadata embedded in the papers themselves, which is well known and researched~\cite{Byers2004Information,Auraetal2006Scanning,Castiglioneetal2007Taking,Alonsoetal2009Disclosing,Garfinkel2013Leaking,Fengetal2018A,Adhataraoetal2021Exploitation}.
To the larger public, the risks specific to publishing sources of preprints were first anecdotally mentioned on Twitter/X~\cite{QuantPhComments2018Overheard}.
However, the underlying issue gained wider attention when a paper on \mbox{GPT-4} purposefully omitted details concerning its toxicity probability and how it complies with intellectual property laws~\cite{DV25591069650762023Sparks}.

Still, despite \arxiv{} maintaining its own build pipeline and warning about corresponding issues~\cite{arXiv2005Requiring}, proactive mitigation of those issues at scale seems to be lacking.
While \arxiv{} endorses~\cite{arXiv2005Requiring} Google's arxiv\_latex\_cleaner (ALC)~\cite{Google-Research2019arxivlatexcleaner,Pont-Tuset2019arXiv}---a tool claiming to ``easily clean the \LaTeX{} code of [..] paper[s]''---it does not enforce any actions on the submitting authors.
Similarly, while anecdotal evidence and a few academic works~\cite{Cormodeetal2012Scienceography:,Cormodeetal2013Studying} hint at the issue of unintentionally distributing comments and other data through \LaTeX{} source files, a detailed analysis of the prevalence and scale of security problems associated with the distribution of papers' sources is lacking.
However, the lack of understanding these problems challenges us in assessing respective risks, notifying affected users, and taking informed countermeasures.

Tackling this research gap, we set out to quantify the prevalence of ``hidden'' information, including sensitive data, within \emph{all} \LaTeX{} sources published by \arxiv{}.
In particular, we consider the availability of unused (dangling) files, embedded metadata, and compilation-irrelevant content (\eg{} comments in \LaTeX{} files) to get a holistic picture.
Indeed, our findings show that ``hidden'' information within \arxiv{} is a widespread issue:
\qty{88}{\percent} of papers with \LaTeX{} sources are affected by at least one type of ``hidden'' information, making information available for everyone to ``discover'' and potentially exploit.
Our analysis reveals substantial amounts of content that were either not meant for distribution or should not be publicly distributed, including reviewer comments, author discussions, survey data with personally identifiable information (PII), as well as links to internal documents and reports without any restrictive access control.
Perhaps surprisingly, papers related to top-tier security conferences contain significantly more ``hidden'' information, despite this community's expectable awareness of such issues.
Overall, the prevalence of issues concerning sensitive information prompted us to initiate two responsible disclosure campaigns (\cf{} \cref{app:sec:ethics}).

\textbf{Contributions.}
To provide a profound understanding of the prevalence and severity of ``hidden'' information within \arxiv{} submissions, better assess corresponding risks, and lay the foundation for mitigations, our main contributions are:
\begin{itemize}[topsep=0em,leftmargin=1em]
	\item We present a methodology to reliably detect irrelevant content in \LaTeX{} sources and identify unused files.
	Unlike current mitigation approaches that solely rely on heuristics, our approach utilizes an abstract syntax tree and \texttt{pdflatex}'s recorder option for precise identification of potentially critical ``hidden'' information (\cref{sec:methodology}).
	\item We conduct the first large-scale analysis of ``hidden'' information within preprint sources based on a comprehensive dataset of \LaTeX{} sources at \arxiv{} (\cref{sec:dataset}), covering \emph{all} \qty{2.7}{\million} submissions with source files.
	\item We reveal that the majority (\qty{88}{\percent}) of papers feature (unnecessary) information within their \arxiv{} submission, with striking findings such as secrets (API tokens, private keys, passwords), sensitive content (links to
internal documents, disclosure of FTP and SSH servers), and uncommented text fragments and discussions among coauthors
	 (\cref{sec:analysis}).
	\item To address the identified serious and prevalent issues, we develop and release \alcpp{}---an improved \LaTeX{} cleaner that reliably prevents ``hidden'' information distribution,
	mitigating \qty{90}{\percent} of ``hidden'' information issues
	(\cref{sec:latexcleaner}).
\end{itemize}

\textbf{Open Science Statement.}
We are committed to open science and thus publish aggregated data of our evaluation~\cite{Pennekampetal2026Artifact:}.
We also release our \LaTeX{} source file cleaner---\alcpp{}---for public use~\cite{Pennekampetal2026ALC-NG}.
In compliance with our ethical considerations (\cref{app:sec:ethics}), we refrain from publishing our analysis code.

\section{Preprint Repositories and \arxiv{}.org}
\label{sec:preprint}

In academia, a preprint corresponds to a scientific paper that is distributed prior to peer review or official publication~\cite{McKiernan2000arXiv.org:,Moshontzetal2021A,Xieetal2021Is}.
While they can be shared any time, around \qty{40}{\%} of preprints get (finally) formally published~\cite{Xieetal2021Is}.
Independent projects (see DOAPR~\cite{COARetal2022Directory}, the directory of open access preprint repositories, for an overview) collect preprints on specific domains, \eg{} bioRxiv and medRxiv, or for specific purposes, establishing different preprint repositories.
Some, such as SSRN or TechRxiv, are even operated and/or endorsed by academic publishers.
In 2021, Xie et al.~\cite{Xieetal2021Is} studied the use of preprint repositories, particularly noting the popularity of \arxiv{}~\cite{Cornell-Tech1991arXiv.org} and identifying a substantial increase in usage in computer science over the past few years.

Established in 1991 to consolidate the then prevalent preprint sharing via email in physics research~\cite{McKiernan2000arXiv.org:}, \arxiv{}~\cite{Cornell-Tech1991arXiv.org} has evolved into a free, open-access preprint repository, hosting more than \qty{2.9}{\million} research papers~\cite{arXiv2014Monthly-Submissions} of different domains~\cite{arXiv2024arXiv}, including natural sciences, computer science, quantitative finance, electrical engineering, and economics.
The importance of preprints has been established decades ago~\cite{Brown1999Information,Manuel2001The}, and the observed growth is domain-independent, \ie{} it also holds for computer security~\cite{Linetal2020How}.
Since its inception, both the number of submissions and downloads have been increasing continuously~\cite{arXiv2024arXiv,arXiv2014Monthly-Downloads,arXiv2014Monthly-Submissions}.

The general submission process at \arxiv{} works as follows.
After writing the paper, an author creates a submission by uploading the \LaTeX{} sources of the paper.
\arxiv{} then compiles the sources and conducts some basic checks to verify whether the paper is eligible for inclusion.
This process is not a proper peer review, but rather a means for maintaining quality within the repository~\cite{Ginsparg2014ArXiv,arXiv2024arXiv}.
Upon successful vetting, the paper is publicly announced, and a \emph{permanent, versioned record} is created.
After the initial announcement, authors can decide to upload a new, improved (corrected) \emph{version} that amends the original record or withdraw a version for all kinds of reasons~\cite{Raoetal2024WithdrarXiv:}.
In the latter case, \arxiv{} still retains the original submission but adds a dedicated notice of withdrawal.

For different reasons~\cite{arXiv2005Requiring}, including accessibility and improved long-term maintainability, \arxiv{} requires authors to submit \LaTeX{} sources if available.
Accordingly, more than \qty{90}{\percent} of submissions are distributed with their sources~\cite{Brinnetal2022A}.

\section{Related Work}
\label{sec:relatedwork}

Research analyzing \arxiv{} has largely focused on non-security aspects, such as
\begin{enumerate*}[label=(\roman*)]
    \item studying how academic papers are written using \LaTeX{}~\cite{Cormodeetal2012Scienceography:,Cormodeetal2013Studying},
	\item understanding the impact of \arxiv{} submissions~\cite{Lariviereetal2014arXiv} and the influence of preprints on citations~\cite{Kurtzetal2005The,Davisetal2007Does,Moed2007The,Haqueetal2009Positional,Haqueetal2010Last,Lariviereetal2014arXiv},
	\item conducting analyses of authorship and publication forms~\cite{Clementetal2019On,Saieretal2020unarXive:,Bianchi2025Text,Hepler2025Modular,Peietal2025Hidden,Yangetal2025Writing},
	\item detecting plagiarism~\cite{Sorokinaetal2006Plagiarism}, text reuse~\cite{Citronetal2015Patterns}, and AI-generated texts~\cite{Akram2024Quantitative},
	\item collecting reasons for withdrawing preprints~\cite{Raoetal2024WithdrarXiv:},
	\item training and evaluating language models~\cite{Daweietal2020A,Ginevetal2020Scientific,Scharpfetal2020Classification}, and
	\item reconstructing \LaTeX{} sources~\cite{Safnuketal2018Reconstructing}, analyzing document layouts~\cite{Lietal2020DocBank:}, or extracting content from \LaTeX{}-compiled PDFs~\cite{Meuschkeetal2023A}.
\end{enumerate*}

However, despite enabling all these diverse research streams, publishing sources of \LaTeX{} documents inevitably carries the risk of unintended information disclosure~\cite{arXiv2005Requiring}.
Indeed, related work analyzing source code in GitHub repositories~\cite{Sinhaetal2015Detecting,Melietal2019How,Fengetal2022Automated,Basaketal2023SecretBench:,Basaketal2023A}, Docker containers~\cite{Dahlmannsetal2023Secrets}, and mobile apps~\cite{Schmidtetal2025Leaky} has identified a substantial risk of disclosing information such as secret API keys, private keys, or passwords.
Consequently, different tools to sanitize \LaTeX{} documents before publishing them on \arxiv{} have been developed~\cite{arXiv2005Requiring,Hughes2012latexindent.pl,Elsa-Lab2019arXiv,Google-Research2019arxivlatexcleaner,Stutz2022Submission}.
While we refer to \cref{subsec:latexcleaner:comparison} for a detailed survey of those tools and their capabilities in accurately sanitizing \LaTeX{} documents, anecdotal evidence of (remaining) critical content in sources of \arxiv{} submissions~\cite{QuantPhComments2018Overheard,DV25591069650762023Sparks,Cormodeetal2012Scienceography:,Cormodeetal2013Studying} suggests that these tools are either not widely applied or ineffective in removing (all) critical content.

Looking beyond source files, information leakage resulting from published documents such as PDF files~\cite{Auraetal2006Scanning,Garfinkel2013Leaking,Fengetal2018A,Adhataraoetal2021Exploitation}, images~\cite{Gouertetal2022Dirty,Nguyenetal2024MetaLeak:}, and other formats~\cite{Byers2004Information,Castiglioneetal2007Taking,Alonsoetal2009Disclosing} has been analyzed, uncovering several common issues, including GPS locations where images were taken, author names, and other (partially) private information unintentionally stored primarily in metadata fields.
To our understanding, these issues of ``hidden'' information in published documents are well-researched and well-understood, with practical tools for metadata removal being available~\cite{Harvey2003ExifTool,Berkenbilt2005qpdf,Voisin2018mat2}.
Further work has focused on detecting malicious PDFs~\cite{Castiglioneetal2010Security,Stevens2011Malicious,Smutzetal2012Malicious}, which is only partly applicable to \arxiv{}-published papers given its service-provided build pipeline, or the identification of privacy issues within PDF-based publication processes~\cite{Adhataraoetal2021How}.

Consequently, research is well aware that ``hidden'' information in documents, including metadata, can be a threat as it may unintentionally reveal sensitive details, \eg{} GPS locations in images or author details in rich text documents.
At the same time, we have \arxiv{}---the largest and most popular preprint service---which, in contrast to most other preprint services, requires and publicly distributes author-submitted \LaTeX{} source files.
Despite selective attention on the Web~\cite{QuantPhComments2018Overheard,DV25591069650762023Sparks} and in academia~\cite{Cormodeetal2012Scienceography:,Cormodeetal2013Studying} as well as warnings issued by \arxiv{}~\cite{arXiv2005Requiring}, research has not yet assessed the prevalence of this phenomenon at scale.
Concurrent to our work, two preprints with specific foci---a LLM-driven security audit of source files~\cite{Dubniczkyetal2025You} and assessing the content of nonrequired files within sources ~\cite{Apruzzeseetal2026X-raying}---also study \LaTeX{} source files, albeit at significantly smaller scale.
Still, they highlight and emphasize the relevance of our research.

\section{Methodology For Analyzing \LaTeX{} Sources}
\label{sec:methodology}

The goal of our work is to discover, quantify, and analyze the distribution of ``hidden'' information at the popular preprint service \arxiv{}.
These findings also serve as a foundation for mitigation strategies to prevent the inclusion of ``hidden'' information.
In line with research in other contexts (\cref{sec:relatedwork}), we identify three key dimensions of ``hidden'' information in \arxiv{} submissions:
\begin{enumerate*}[label=\nostep{\arabic*}{}]
	\item \LaTeX{} sources bundled with dangling \emph{files},
	\item embedded \emph{metadata}, and
	\item \emph{content/text} irrelevant to the build process.
\end{enumerate*}

In the following, we present our methodology to comprehensively uncover ``hidden'' information across these three dimensions with high accuracy to lay the foundation for analyzing \arxiv{} submissions and their source files at scale.

\textbf{Dimensions of ``Hidden'' Information.}
Our methodology for discovering ``hidden'' information aligns with the three dimensions along which such information can occur.

\nostep{1}{} \emph{File Level.}
At the file level, we classify source files as either required or dangling files.
\emph{Required} files are strictly needed to compile the \LaTeX{} sources into a PDF document, while \emph{dangling} files are not used during the build process and thus do not need to be part of the served sources.

We align our detection of dangling files with \arxiv{}'s long-used submission system (version 1.0).
That is, we mark all \LaTeX{} root files with \texttt{documentclass} or \texttt{documentstyle} as required and then recursively mark all additionally included files as required.
We further support \texttt{00README}(\texttt{.XXX}) files~\cite{arXiv2010The}, present in \qty{8.8}{\percent} of all \arxiv{} submissions with sources, in which authors may optionally provide additional information to the submission system, \eg{} to override \arxiv{}'s default inclusion behavior for files.

\nostep{2}{} \emph{Metadata Level.}
Most file types further support the inclusion of metadata to add supplementary information and provenance data.
Since related work (\cf{} \cref{sec:relatedwork}) has shown that metadata may leak sensitive information~\cite{Spiegel2017Are}, we rigorously analyze all source files using exiftool~\cite{Harvey2003ExifTool}.

\nostep{3}{} \emph{Content/Text Level.}
Besides content within required files that is needed to produce the PDF document, source files on \arxiv{} can also contain \emph{irrelevant} content with no influence on the compiled PDF.
In this regard, \LaTeX{} source files (beyond other file types) are especially relevant given their intentional support for author-supplied comments~\cite{Cormodeetal2012Scienceography:}---a common practice in (academic) writing.

To identify irrelevant content in \LaTeX{} sources, our methodology relies on an abstract syntax tree representation of said \LaTeX{} document using a tree-sitter~\cite{tree-sitter2018Tree-sitter} grammar.
Harnessing this structured representation of \LaTeX{} documents, we can reliably parse and extract irrelevant content, \eg{} comments, by specifying the different ways irrelevant content can be embedded in \LaTeX{} sources.
This approach fundamentally differs from the current state of the art in sanitization tools (\cf{} \cref{subsec:latexcleaner:comparison}), which rely on heuristic approaches based on regexes.
To ensure comprehensive coverage of the different concepts of embedding comments, we survey the treatment of irrelevant information by six state-of-the-art \LaTeX{} sanitization tools (\cf{} \cref{subsec:latexcleaner:comparison}) and supplement these findings with our own experience from authoring more than one hundred academic papers with various research groups across three continents.

\textbf{Syntax of Embedded Comments.}
Using this approach, we derive four widely-used and commonly-known concepts for embedding comments in \LaTeX{} documents that we employ to detect irrelevant content:
\begin{enumerate*}[label=\nostep{\Alph*}{}]
	\item standard \LaTeX{} line comments preluded by \texttt{\%} or wrapped in a \texttt{comment} environment,
	\item any characters outside of \LaTeX{}'s \texttt{document} environment,
	\item non-taken branches of (custom) \texttt{if} statements, and
	\item ignored arguments of (\texttt{re})\texttt{newcommand}, \texttt{providecommand}, \texttt{Declare*}, or \texttt{def} constructions.
\end{enumerate*}
Complementing the content-level dimension, we further consider all content in dangling files as irrelevant.

Given \LaTeX{}'s expressiveness, we may theoretically miss other irrelevant content that does not match the syntax described above; for example, \texttt{csname} constructions to insert content or author-implemented conditionals.
However, our investigation into the usage frequency of custom commands did not reveal meaningful findings.
Thus, such practices seem rare within \arxiv{} sources, \ie{} they are not used at scale.

\textbf{Handling Multiple Versions of a Submission.}
When analyzing \arxiv{} papers, multiple \emph{versions} (\ie{} updates) of a single paper must be considered (\cf{} \cref{sec:preprint}).
Issues present in one version may not be part of another version of the same paper (either the issue gets introduced in a later version or removed during an update).
To avoid overcounting ``hidden'' information, we thus perform a worst-case aggregation over versions.
That is, we consider an issue to be present for a given paper if we discover it in \emph{at least} one version, not necessarily the most recent one.
As the sources of older versions remain accessible for anyone, this approach is reasonable and accurately quantifies information leakage.

\textbf{Validity.}
By relying on a structured parsing approach based on an abstract syntax tree (instead of relying on heuristics, \eg{} based on regular expressions), our approach intentionally constitutes an accurate lower bound of detected irrelevant content.
Furthermore, as our coverage of the different concepts of embedding comments relies on best practices from existing sanitization tools and extensive authoring experience, we are convinced that we capture the vast majority of irrelevant content.
While \LaTeX{}, as a Turing-complete language, theoretically offers additional custom approaches for embedding comments, as part of our analysis, we did not discover such peculiarities at scale and thus estimate their prevalence to be negligible in practice. %
As such, our assessment methodology
\begin{enumerate*}[label=(\alph*)]
	\item reasonably minimizes false negatives (undetected irrelevant content), and
	\item yields almost no false positives (relevant content falsely flagged as irrelevant).
\end{enumerate*}
Alongside turning our comprehensive detection methodology into a next-generation sanitization approach (\cf{} \cref{subsec:latexcleaner:mitigation}), we will show that our abstract syntax tree-based approach outperforms all existing sanitization tools in simultaneously minimizing false negatives and false positives while not visually changing compiled documents.

To the best of our knowledge, we are the first to apply a methodological approach to assess the prevalence of ``hidden'' information in \LaTeX{} source files at scale.
Our methodology ensures an assessment that minimizes false positives, \ie{} we do not label content as irrelevant if it is strictly needed when compiling the PDF from source.
In contrast, although covering all prevalent approaches to embed irrelevant content, our approach does not guarantee completeness; rare custom cases might be missed.
Consequently, our results constitute a reliable lower bound of discovered ``hidden'' information.

\section{Dataset Covering All \arxiv{} \LaTeX{} Sources}
\label{sec:dataset}

For our analysis, we have curated a dataset covering the source files of \emph{all} \arxiv{} submissions from its launch in 1991 up to and including 12/2025.
For 1991--10/2020, we retrieved the \url{archive.org} collection~\cite{bnewbold2017arXiv.org}.
To retrieve the remaining months and fill any gaps, we relied on the official arxiv-bulk dataset access over Amazon S3 storage~\cite{arXiv2010Full}.
Since these sources only serve one version of a paper, we further sampled full version histories using the official \arxiv{} API while complying with their rate limits (\cf{} \cref{app:sec:ethics}).

After retrieval, we validated the completeness of our dataset:
\begin{enumerate*}[label=(\roman*)]
	\item we iterated over the paper ID space to check for gaps (some IDs are not associated with a submission),
	\item we marked all withdrawn versions and papers (similar to WithdrarXiv~\cite{Raoetal2024WithdrarXiv:}), and
	\item we verified that submissions without source files in our dataset are indeed without sources.
\end{enumerate*}
For the latter two steps, we were in contact with \arxiv{}.

\begin{figure}[t]
	\centering
	\includegraphics[width=\linewidth]{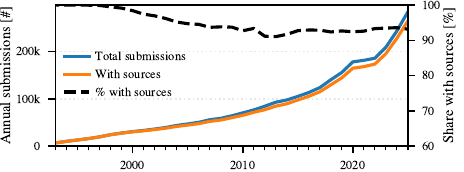}
	\vspace{-2em}
	\caption{%
		Annual volume of \arxiv{} submissions with and without sources.
	}
	\vspace{-1em}
	\label{fig:dataset}
\end{figure}

In total, our dataset covers \qty{2.7}{\million} papers with source files, which corresponds to \qty{93}{\percent} of all papers on \arxiv{}.
\cref{fig:dataset} shows their distribution over the years, with \num{2013} representing the relative lower bound (\qty{91}{\percent}) and \num{1991} the upper bound (\qty{100}{\percent}).
Across the papers with source files, we sampled \qty{1.1}{\million} papers with more than one version with source files, accumulating to a total of \qty{4.3}{\million} versions.

\section{Analysis of \arxiv{} Source Files at Scale}
\label{sec:analysis}

Our dataset of all \qty{2.7}{\million} submissions with available source files published on \arxiv{} (by the end of 2025) enables us to discover and analyze ``hidden'' information within sources made available by \arxiv{} and thus derive a comprehensive picture of the prevalence of unintentional information disclosure in source files of scientific preprints.
To this end, we first provide a high-level overview of our findings, summarizing frequent types of ``hidden'' information and reporting changes over the years, \arxiv{} categories, and submitted paper versions (\cref{subsec:analysis:bigpicture}).
Subsequently, we attribute our findings to the different dimensions of the problem (\cf{} \cref{sec:methodology}) and sequentially discuss them in greater detail (\cref{subsec:analysis:individual}).

We focus on assessing the prevalence of ``hidden'' information and grasping its sensitivity.
Consequently, the implications of content-specific changes over time, \eg{} the removal or addition of authors, the effects of changing templates, or the creation of citation graphs, which related work already covered partly (\cf{} \cref{sec:relatedwork})---albeit at a comparatively small scale---are not in our scope. %

\subsection{A Birds-Eye View at ``Hidden'' Information}
\label{subsec:analysis:bigpicture}

We begin our analysis by examining the rate of ``hidden'' information at the file, metadata, and content/text levels.

\textbf{Guiding Question.}
\emph{%
How prevalent are the three dimensions of ``hidden'' information in source files of preprints?
}

To provide insights into this guiding question, we filter for interesting content (\cref{subsubsec:analysis:bigpicture:content}), quantify frequent issues (\cref{subsubsec:analysis:bigpicture:overview}), and analyze differences across submission categories, versions, and years (\cref{subsubsec:analysis:bigpicture:differences}).

\subsubsection{Filtering For ``Interesting'' Content}
\label{subsubsec:analysis:bigpicture:content}

\begin{figure}[t]
	\centering
	\includegraphics[width=\linewidth]{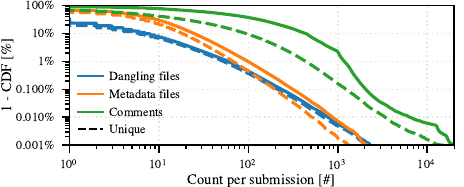}
	\vspace{-2em}
	\caption{%
		Complementary cumulative distribution functions (1-CDFs) of total and unique ``hidden'' information per \arxiv{} submission across the three dimensions: dangling files, metadata key-value pairs, and \LaTeX{} comments.
	}
	\label{fig:information-cdf}
	\vspace{-1em}
\end{figure}

\begin{figure*}[t]
	\centering
	\includegraphics[width=\linewidth]{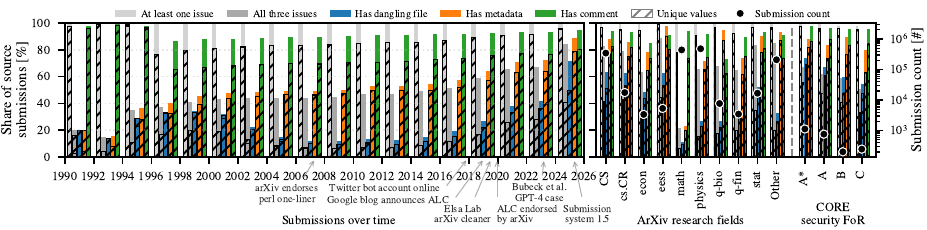}
	\vspace{-2em}
	\caption{%
		Development of the prevalence of ``hidden'' information over time shows an upward trend (left) as well as across aggregated \arxiv{} categories and the CORE~\cite{Computing-Research--Education2013ICORE} field of research (FoR) 4604 ``cybersecurity and privacy'' (right), revealing minor differences across research fields and publication venues.
	}
	\label{fig:issues-overview}
	\vspace{-1em}
\end{figure*}

After processing the entire dataset of \arxiv{} sources, we obtain a massive volume of ``hidden'' information for analysis across all considered dimensions, including \qty{12}{\million} dangling files, \qty{144}{\million} metadata key-value pairs, and \qty{644}{\million} comments.
\cref{fig:information-cdf} shows their distribution across submissions.
However, a substantial fraction of this massive volume of ``hidden'' information repeats across (many) submissions (e.g., unused template files or comments in templates) and thus likely does not constitute potentially unintentionally-disclosed information.

\textbf{``Interesting'' Content.}
To instead focus only on potential cases of unintentional information disclosure, we define ``interesting'' content as any content that is (almost) \emph{unique} to an individual submission, following the intuition that ``hidden'' information occurring only in one submission likely was added by the authors of this submission (and not inherited, \eg from templates).
More specifically, we filter out any dangling file, metadata value, or comment that we find in more than two submissions (to account for rare duplicates by chance or reuse by the same authors).

In \cref{fig:information-cdf}, we visualize complementary cumulative distribution functions (1-CDFs), detailing the number of unique and total information for the three dimensions.

\textbf{File Level.}
Out of \qty{50}{\million} total files, we find \qty{45}{\million} (\qty{91}{\percent}) to be unique (occurring at most twice) based on SHA256 fingerprints (\qty{44}{\million}/\qty{88}{\percent} for strict uniqueness).
This methodology filters out \qty{181}{\thousand} distinct files that we, on average, find in \num{23.6} different submissions (median=\num{3}, max=\qty{57}{\thousand}).
\LaTeX{} templates (.tex/.sty/.cls/.bst) account for \qty{66}{\percent} of these files, and \qty{15}{\percent} correspond to images or PDFs.

\textbf{Metadata Level.}
Metadata either stems from file-level modification timestamps (\qty{90}{\percent} of submissions contain such data, \ie{} were uploaded as an archive preserving this information) or from information embedded in images, PDFs, XML, and other file types that can be extracted using exiftool~\cite{Harvey2003ExifTool} (\qty{93}{\percent} of submissions).
Out of the timestamps, \qty{98}{\percent} are unique, indicating that they are not artifacts of packing submissions for upload to \arxiv{} (per submission, \qty{94}{\percent} of files have unique timestamps).
Interestingly, \qty{84}{\percent} of filtered timestamps coincide with filtered files, confirming a common origin, such as from a ZIP file of a \LaTeX{} template.
For metadata extracted using exiftool, we count a total of \qty{111}{\million} key-value pairs, of which \qty{38}{\percent} are unique.
Still, non-unique metadata values can carry relevant information, \eg{} common software producer names or version numbers.

\textbf{Content/Text Level.}
We identify \qty{644}{\million} independent \LaTeX{} comments in total; the majority of comments (\qty{93}{\percent}) are part of required files.
Comments are present in \qty{95}{\percent} of all considered \arxiv{} submissions.
After filtering for unique comments, we obtain \qty{114}{\million} ``interesting'' comments for analysis that span across \qty{75}{\percent} of the considered submissions.

Our analysis not only reveals a large amount of ``hidden'' information across all three dimensions, but also shows that the majority of this information is sufficiently unique to likely originate from authors directly.
Consequently, our analysis in the following focuses on precisely this ``interesting'' data, as it is substantially more likely to contain sensitive information than repeatedly occurring data (\eg{} template files, common producer metadata values, or comments in template files).

\subsubsection{Quantifying Frequent Issues}
\label{subsubsec:analysis:bigpicture:overview}

To gain a better understanding of the frequency of specific aspects of ``hidden'' information, we compare the prevalence and interplay of its three dimensions.
As we visualize in \cref{fig:issues-overview}, most \arxiv{} submissions contain ``interesting'' comments (\qty{75}{\percent}).
In contrast, fewer submissions contain unique metadata (\qty{56}{\percent}) or unique dangling files (\qty{25}{\percent}).
When looking at submissions that contain ``hidden'' information from more than one dimension, we observe that comments and metadata occur together most frequently (\qty{30}{\percent}).
\qty{18}{\percent} of submissions even contain ``hidden'' information from all three dimensions;
most (\qty{88}{\percent}) contain ``hidden'' information in at least one of the dimensions.

\subsubsection{Differences in Years, Research Fields, and Versions}
\label{subsubsec:analysis:bigpicture:differences}

To better understand potential factors influencing the prevalence of these discoveries, we analyze the prevalence of ``hidden'' information for
\begin{enumerate*}[label=(\roman*)]
	\item changes over time,
	\item differences across disciplines, and
	\item observed variance between different versions of the same submission hosted at \arxiv{}.
\end{enumerate*}

\textbf{Years.}
As we show in \cref{fig:issues-overview} (left side), the prevalence of ``hidden'' information in \arxiv{} source files slowly increases over time across the dimensions.
This observation particularly holds for the prevalence of dangling files (starting in 2013), and holds for all dimensions since 2017.
Perhaps surprisingly, events such as the release or recommendation of sanitization tools or (negative) publicity surrounding ``hidden'' information have no noticeable impact on the observed trends.

\textbf{Research Fields.}
As a multi-disciplinary preprint repository, \arxiv{} is structured into \num{38} categories.
To understand whether the prevalence of ``hidden'' information shows research domain-specific differences, we aggregate these categories into nine research fields for comparison.
As we detail in \cref{fig:issues-overview} (right side), some disciplines, \eg{} Math or Physics, feature substantially fewer ``interesting'' dangling files, potentially because their submissions overall consist of fewer files.
Given their high share of submissions (black dots in \cref{fig:issues-overview}), they have a substantial impact on the overall results that we report.
Thus, covering different research fields when quantifying ``hidden'' information in \arxiv{} source files is important to gain a comprehensive picture.

In addition to this high-level comparison, we further assess whether computer science researchers, in general, and security researchers (A\textsuperscript{*}-/A-/B-/C-ranked conferences), in particular, exhibit any differences in their behavior (\Cref{app:sec:security} details how we match \arxiv{} submissions to published papers by title using dblp~\cite{Schloss-Dagstuhl1993dblp:}).
A two-sided Mann-Whitney U test confirms significant differences between all dimensions shown in \cref{fig:issues-overview} when comparing CS (with cs.CR, cryptography and security) against the rest, as well as between cs.CR and the rest of computer science ($p<0.0001$ in all cases):
There is more ``hidden'' information in computer science papers compared to other disciplines, and even more in security-related papers.
Interestingly, the same observation holds when comparing A\textsuperscript{*} security papers with A ones ($p<0.0001$).
Again, higher-ranked venues feature a significantly larger share of ``hidden'' information.
Normalizing issues by the \LaTeX{} file size (in bytes) to account for length differences does not change this observation.
A potential explanation could rather be that computer scientists (and authors of A\textsuperscript{*} papers) have more elaborate source files, while not necessarily paying more attention to ``hidden'' information.
On a positive note, out of \num{1974} dblp-matched security papers also submitted to \arxiv{}, we discover \num{243} submissions without any dangling files or other irrelevant content.
When also considering the absence of metadata, we identify \num{70} ``perfectly clean'' security papers.

\begin{figure}[t]
	\centering
	\includegraphics[width=\linewidth]{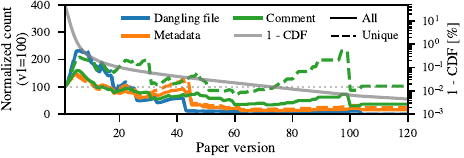}
	\vspace{-2em}
	\caption{%
		Subsequent versions rather add unique data over the first version.
		Only in the long tail, few papers with many versions skew this trend.
	}
	\vspace{-1.5em}
	\label{fig:versions-cdf}
\end{figure}

\textbf{Versions.}
To support the evolution of a preprint over time, \arxiv{} allows superseding a submission with a new \emph{version} (\cf{} \cref{sec:preprint}).
While tools for comparing the PDF content of papers over versions exist~\cite{Emken2020comparxiv,Nanavatietal2021ArxivDiff}, the impact of new (updated) versions on ``hidden'' information remains unknown so far.
As we illustrate in \cref{fig:versions-cdf}, many authors update their \arxiv{}-hosted papers over time, even repeatedly, or withdraw them~\cite{Raoetal2024WithdrarXiv:}, thereby marking them as obsolete (their sources remain publicly available).
Intuitively, one reason to submit a new version of a paper could be an attempt to fix or to remove ``hidden'' information that was accidentally published.
Indeed, individual examples such as a submission initially containing many valid OpenID tokens in dangling files that were later removed with an updated version confirm this suspicion (although this practice is an ineffective mitigation, as the source files of old versions remain accessible). %
However, at scale, we do not see any consistent indication that authors in general reduce the amount of ``hidden'' information when updating their preprints with a new version, when comparing the prevalence of ``hidden'' information between versions in \cref{fig:versions-cdf} (in the long tail, the low number of papers with many, \num{\ge20}, versions prevents deriving well-founded conclusions).

\textbf{Takeaway.}
\emph{%
All three dimensions of ``hidden'' information are prevalent issues, collectively affecting \qty{88}{\percent} of \arxiv{} source-code submissions across years, research fields, and submission versions, especially when focusing on ``interesting'', \ie{} seldomly occurring, content.
}

\subsection{Individually Looking at ``Hidden'' Secrets}
\label{subsec:analysis:individual}

While our initial analysis already establishes the prevalence of ``hidden'' information in \arxiv{} submissions, so far we only know that the majority (\qty{88}{\percent}) of submissions contain interesting, \ie{} unique ``hidden'' content.
We now take a closer look at the individual pieces of ``hidden'' information.

\textbf{Guiding Questions.}
\emph{%
Which individual cases of ``hidden'' information can we discover?
How sensitive are these cases?}

To answer these questions, we examine the three dimensions of ``hidden'' information (file, metadata, and content/text) individually.
To this end, we focus on identifying potentially sensitive information disclosure on a \emph{per-submission} basis, \ie{} if we identify an issue in multiple versions of the same paper, we only count and report it once. %

Where possible, we follow the approach of Leon et al.~\cite{Leonetal2013What} to label findings as high (\impH), medium (\impM), or low (\impL) impact based on the \emph{potential} severity of them being publicly available.
However, this labeling does not imply that all reported findings indeed have the labeled impact.

\subsubsection{\step{1}{}: Bulk of Dangling Files}
\label{subsubsec:analysis:individual:dangling}

\begin{figure}[t]
	\centering
	\includegraphics[width=\linewidth]{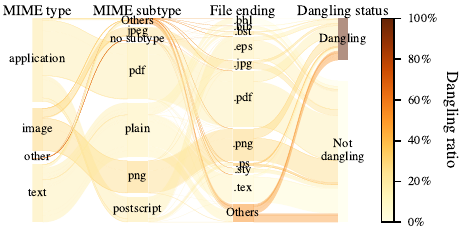}
	\vspace{-2em}
	\caption{%
		\arxiv{} source files contain a broad range of file types.
		Some types, \eg{} images, are more frequently dangling files than others, \eg{} \TeX{} files.
	}
	\vspace{-1.25em}
	\label{fig:danglingfiles-sankey}
\end{figure}

Overall, we discover a total of \qty{12.0}{\million} dangling files.
Digging into file and MIME types, we observe the distribution shown in \cref{fig:danglingfiles-sankey} with a similar distribution across dangling and non-dangling files for frequent file types and a long-tail of rare file types that are predominantly dangling files, including Readme files (\qty{3.4}{\percent} of submissions, \impL), Python code (\qty{0.21}{\percent}, \impL), shell scripts (\qty{0.21}{\percent}, \impL) and other executable files (\qty{0.041}{\percent}, \impL).
We also find research-related data in CSV (\qty{0.20}{\percent}, \impM) and SQLite form (\qty{0.002}{\percent}, \impM).
\qty{0.007}{\percent} of submissions (intentionally) contain ``dangling'' files within the \texttt{anc/} directory~\cite{arXiv2010Ancillary} (\cf{} \cref{subsec:latexcleaner:mitigation}), which we thus do not interpret as dangling.

In general, we cannot reliably assess whether dangling files have been intentionally submitted or not.
While \qty{173}{\thousand} dangling files carry a path such as \texttt{supplementary} or \texttt{additional\_data}, likely indicating intentional submission, \qty{256}{\thousand} files correspond to directories that contain \texttt{backup}, \texttt{old}, or similar, instead indicating unintentional submission.
Aggravatingly, \num{40} submissions contain hidden \texttt{.nfs} files that are created by NFS filesystems immediately after removal (\impM), and \num{74} contain complete Git repositories, including editing histories that we consider unintentionally submitted (\impM).
Furthermore, we discover configuration files that can carry sensitive information such as API keys or database connection strings in \num{2774} submissions (\impH, we detail those findings in \cref{subsubsec:analysis:individual:regexable}).

Bibliography (\texttt{bib}) files are a special case of dangling files, as they are redundant with intermediate \texttt{bbl} files that are sufficient for compilation.
While \texttt{bbl} files comprise only required bibliography entries, we find that \qty{93}{\percent} of \texttt{bib} files contain entries that are not referenced in the compiled document.
Within these dangling \texttt{bib} files, we identify two frequent patterns:
\begin{enumerate*}[label=(\alph*)]
	\item \qty{8.4}{\percent} of submissions reuse bibliographies from prior submissions as is (\impL), and
	\item \qty{5.5}{\percent} of submissions contain large bibliographies from online repositories, such as ACL Anthology~\cite{ACL-Anthology2013Full}, with more than \num{500} entries (\impL).
\end{enumerate*}
We find that especially custom-generated and historically-grown bibliographies reveal more information, including note fields (\qty{49}{\percent} submissions, \impM) and anecdotal citation keys, for example, indicating that authors were scooped by related work (\eg{} \texttt{oldpaperthatscoopedus}, \impM). %

Conference-specific templates are another angle that may reveal information such as submission histories (\impL).
Exemplarily analyzing a small, non-representative sample of \num{115} submissions with dangling \texttt{usenix.sty} files where we could match paper titles against dblp records, we find \qty{77}{\percent} of these submissions at venues not using USENIX templates (A\textsuperscript{*}: \qty{61}{\percent}, A: \qty{30}{\percent}, B: \qty{5.2}{\percent}, C: \qty{3.5}{\percent}).
While we cannot certainly claim that these papers were indeed rejected from USENIX-hosted conferences, these examples suggest that tracing submission histories based on \arxiv{} data could be feasible.
We observed and manually validated similar trends for templates of PETS, NeurIPS, CVPR, and ICCV.
When identifying venue-specific templates (.tex/.sty/.cls/.bst) as ``more than \qty{80}{\percent} of template usages occur at the same venue according to dblp'', we count \num{47} venues and \num{561} affected submissions that may be traceable this way (\impL).
This aspect concerns \qty{0.24}{\percent} of the dblp corpus we can study.

Overall, when focusing on files that are unnecessarily submitted to \arxiv{}, we identify both files that in themselves are sensitive, \eg{} \texttt{.nfs} files or Git histories, as well as files that allow us to infer further, less severe contextual information, such as unused template files hinting at submission histories or \texttt{.bib} files revealing collaboration patterns.

\subsubsection{\step{2}{}: Insightful Metadata}
\label{subsubsec:analysis:individual:metadata}

\begin{figure}[t]
	\centering
	\includegraphics[width=\linewidth]{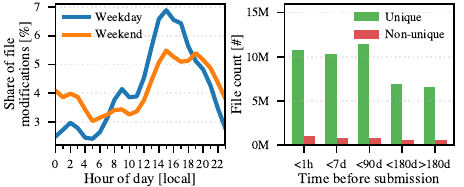}
	\vspace{-2em}
	\caption{%
		File modification timestamps allow for deriving author behavior from \arxiv{} source files, revealing details of the respective authoring process.
	}
	\vspace{-1.25em}
	\label{fig:metadata-timestamps}
\end{figure}

As authors typically upload a compressed archive that includes original timestamps to \arxiv{}, we can analyze these timestamps and other metadata, which allow for deriving potentially sensitive (\impM) author-specific behavior~\cite{Adhataraoetal2021How,21jfx5522022suggestion:}.
\qty{72}{\percent} of submissions carry modification timestamps within one hour (\qty{35}{\percent} within five minutes) before arXiv-recorded submission timestamps.
In \cref{fig:metadata-timestamps}, we detail the extracted modification times, confirming submission date-based patterns observed in prior work~\cite{Haqueetal2010Last} despite extending our analysis to all source files.
For creation times, we observe wide ranges, with \qty{83}{\percent} of submissions indicating creation dates more than \num{180} days before submission, reflecting long-term authoring processes.
Here, we excluded commonly reused templates, as we only found \qty{38}{\percent} of them carrying unique timestamps, whereas most timestamps seem to originate from template providers.

To study file-embedded metadata, we run exiftool~\cite{Harvey2003ExifTool} on all source files.
This method yields \qty{62}{\percent} of non-dangling files and \qty{61}{\percent} of dangling files carrying metadata, most frequently in images (\qty{90}{\percent} of all images) and PDFs (\qty{93}{\percent} of all PDFs).
While embedded images have been studied before~\cite{Garfinkel2013Leaking,Fengetal2018A}, our source-based analysis extends this analysis to also cover metadata of other file types, including dangling files, that have not been part of prior analyses.

Only considering information not accessible from the compiled PDF, we find \qty{11}{\percent} of submissions leaking usernames (\impM), \qty{60}{\percent} software information (\impL), \qty{2.9}{\percent} email addresses (\impM), and \qty{1.4}{\percent} specific hardware (\impL), \ie{} a significant exposure of sensitive authoring information~\cite{Adhataraoetal2021How}.
In line with related work, we also extract metadata that leaks GPS locations in \num{7326} of submissions (\impH), \qty{34}{\percent} of those findings are limited to dangling files.
\num{2238} sources contain multiple distinct coordinates; \num{235} of them have maximum distances of at most \qty{50}{\kilo\meter} (at least \qty{0.3}{\kilo\meter}), which we consider commutable, thus possibly enabling the tracking of authors in the temporal and spatial dimensions.
Randomly sampling ten of them, we indeed find nine submissions that feature both research buildings and residential areas (\impH).

\subsubsection{\steptwo{3a}{}: Noticeable (Regexable) Content}
\label{subsubsec:analysis:individual:regexable}

\begin{figure}[t]
	\centering
	\includegraphics[width=\linewidth]{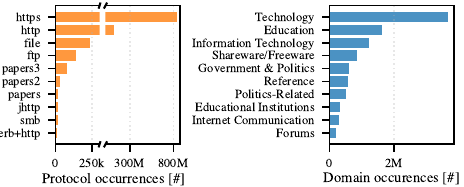}
	\vspace{-2em}
	\caption{%
		Occurrence of URLs in our dataset grouped by protocol (left) and domain category (right) based on Cloudflare's domain classification~\cite{Cloudflare-Inc.2025Domain}.
	}
	\vspace{-1em}
	\label{fig:domains-protocols}
\end{figure}

We now turn our attention to the content/text level within sources that can be efficiently identified through pattern matching.
While regex-based methods require further validation and can yield false positives~\cite{Basaketal2023A}, they provide an efficient mechanism for assessing potentially sensitive content at scale.
We thus employ an extensive collection of regular expressions~\cite{Ahmed2023Secrets} to identify concerning content candidates, including cryptographic material, credentials, API keys, network resources, and other patterns related to potentially sensitive information disclosure.
For credentials and keys, we employ filtering as recommended in recent related work~\cite{Zhouetal2025Hey}, namely a segmented Shannon-Entropy filter ($>$ 3), besides discarding strictly increasing or decreasing character sequences and a known list of patterns (\eg{} \texttt{aaaa}).
We run all patterns against all text files in our dataset.
\cref{tab:appendix:regex-matches} in \cref{app:sec:evalextra} provides a quantitative overview of our findings.
We responsibly disclosed validated critical findings to allow authors to take corrective actions (\cf{} \cref{app:sec:ethics}).

\textbf{Protocol Diversity.}
We extract and analyze all embedded URLs and IP addresses to understand which external resources authors reference and expose through source files.
In total, we identify \qty{127}{\million} unique URLs (\impL) across \qty{905}{\thousand} submissions and \qty{41}{\thousand} IP addresses (\impL) in \qty{19}{\thousand} submissions.
When analyzing the protocol used in URLs, besides the prevalent use of HTTP(S) (\qty{98}{\percent} of URLs), we also discover references to (legacy) protocols (\cref{fig:domains-protocols} left), including FTP servers (\qty{29}{\thousand} submissions, \qty{28}{\percent} online, \qty{23}{\percent} with read access, \impM{}), SSH connection strings (\num{349} submissions, \impL{}), and other network protocols that may inadvertently reveal infrastructure details or serve sensitive content.

\textbf{Web References.}
Leveraging Cloudflare's domain classification records~\cite{Cloudflare-Inc.2025Domain}, we further categorize discovered domains in 10/2025 and present their distribution in \cref{fig:domains-protocols} (right).
Among frequent domains, we find links to cloud storage services, including Google Drive (\num{4272} submissions, \impM), Dropbox (\num{1542}, \impM), and others, as well as collaborative platforms such as Overleaf (\qty{26}{\thousand}, \impM) and Google Docs (\num{3948}, \impM).
Manually accessing a randomly sampled subset of these links reveals that some of the linked services are publicly accessible, \ie{} they might serve sensitive content unintentionally.
Overall, \num{4283} embedded links were---at the time of the conducted classification---labeled as ``file sharing'' by Cloudflare.
Moreover, as classified in 10/2025, we discover \num{587} links related to questionable content (incl.\ CIPA filter, profanity, and pornography, \impL); \num{7} links serve violent content (according to Cloudflare, subject to change).

\textbf{Content Classification.}
Exemplarily looking at the content behind found web references (\cf{} \cref{app:sec:ethics} for ethical considerations), we observe that \num{1119} Google Docs links grant viewing (\impM) or \num{699} editing (\impH) access, with at least \num{200} cases exposing sensitive content (according to manual analysis, \impH).
This content covers commonly-confidential documents from the authoring process (reviews, cover letters, rebuttals, and revision logs), todo lists with both ticked and unticked items, meeting minutes (in one case, including links to Zoom recordings), student assignments, and even a shift schedule, as well as a stock and order list.
For other content types, \eg{} datasets, evaluation results, figures, or plots, we cannot assess the sensitivity without additional context.
We also find several Google Docs links that allow us to transitively reach other web resources not referenced in the source files, resulting in broader information disclosure.
Alarmingly, we also came across \num{18} cases where we obtained access to survey data of study participants (\impH).

\textbf{Security-related.}
Further analyzing irrelevant content for clearly security-critical content, we discover \num{265} API tokens in \num{128} submissions (\qty{26}{\percent} dangling, \impH), \num{4} private keys in \num{4} submissions (\qty{50}{\percent} exclusively in dangling files, \impH), and \num{171} generic passwords in \num{82} submissions (\qty{25}{\percent} dangling, \impH) after manual validation.
Their placement within the source files indicates that the authors likely did not intend to distribute them publicly.
These discoveries align with related work on secret detection in software repositories~\cite{Sinhaetal2015Detecting,Melietal2019How,Basaketal2023SecretBench:} and highlight that the problem extends beyond traditional code-sharing platforms to academic preprint services.

\textbf{Malware Detection.}
Using YARA's established pattern-based approach for detecting malware~\cite{VirusTotal2008YARA}, we attempted to discover malicious (binary) patterns within the source files.
However, upon manual validation, all hits turned out to be false positives.
Thus, this straightforward approach for identifying malicious content did not reveal hazardous risks.

\textbf{``Hidden'' LLM Instructions.}
We also investigated the source files for indicators of problematic authoring activities.
Recent work has highlighted concerns about LLM-generated content being used to influence peer review~\cite{Sugiyamaetal2025Positive,Gelman2025IGNORE,Colluetal2025Publish}, prompting us to search for patterns that resemble prompts or instructions to language models.
Apart from confirming prior findings of instructions within the PDF~\cite{Colluetal2025Publish}, we specifically looked for prompt injection patterns in commented-out sections, \ie{} instructions that may have deliberately been part of the paper before, \eg{} during peer review.
In fact, we manually confirmed nine submissions containing such instructions.
Besides, 537 submissions embed hidden text fragments that match common LLM response disclaimers like ``As an AI language model'' in commented-out sections, which indicates generated content~\cite{Westeretal2024As} (\impL).
However, their actual purpose and impact on the document's text remain unclear without deeper semantic analysis.

\textbf{Other Noticeable Content.}
Similarly, searching for review-indicating keywords such as ``reviewer,'' ``rebuttal,'' or phrases such as ``wrong'' and ``not correct'' yields 3.2\,K submissions containing such terms in commented parts, potentially indicating responses to peer feedback that authors forgot to remove before submission (\impL).
We also discover that 269 submissions actively use \LaTeX{} packages designed for content redaction, such as \texttt{censor} and \texttt{pdfprivacy}, suggesting attempted but ineffective sanitization (\impH).
Further, within dangling files, we encounter LICENSE files in \num{5602} submissions (\impL), which may be incompatible with the submission's license, potentially unintentionally open-sourcing differently-licensed material.
This finding raises questions on intellectual property considerations and whether authors understand the implications of distributing their research under specific open-source licenses through \arxiv{}.

Finally, applying profanity detection patterns~\cite{Ahn2009Offensive/Profane}, we discover offensive language in citation keys, \eg{} \texttt{fuck1} to \texttt{fuck7}, appearing in \num{101} submissions (\impL).
This finding carries implications also for generated PDF documents, as citation key names and reference labels become accessible in PDF metadata when using the hyperref package---which \arxiv{} enabled by default in its original submission system (1.0)~\cite{arXiv2008HyperTeX}.
In this context, we identify \qty{42}{\million} custom---likely irrelevant---bibliography entry fields that deviate from standard \BibTeX{} schemas, mostly containing editorial comments or annotations that were likely not intended for public distribution:
\qty{29}{\million} entries across \qty{150}{\thousand} submissions reveal author comments, local file paths, and behavior, \ie{} creation and modification dates as well as ratings of papers (\impL).

\subsubsection{\steptwo{3b}{}: Author Comment Analysis}
\label{subsubsec:analysis:individual:comments}

Moving beyond noticeable content that can be identified via pattern-matching, we now strive to identify more subtle aspects of ``interesting'' content, \ie unique author comments.

\textbf{Methodology.}
To better understand these individual comments at scale (see \cref{sec:methodology} for what we consider a comment), we task an LLM with classifying the content, building on promising results for text classification and sentiment analysis across diverse domains~\cite{Weietal2023Empirical,Huangetal2025Text,Vajjalaetal2025Text}.
Given the potentially sensitive nature of the content, we utilize a locally-hosted \texttt{Qwen2.5-72B} multi-language model to avoid transmitting any content to third-party services (\cf{} \cref{app:sec:llmusage} for a detailed discussion of our LLM usage considerations).
According to lingua-py~\cite{Stahl2022lingua-py}, \qty{92}{\percent} of comments are in English language, followed by Latin (\qty{2.9}{\percent}, often missclassified math commands), European (\qty{2.5}{\percent}), and Asian (\qty{0.6}{\percent}) languages.
Overall, $>$\qty{96}{\percent} of comments correspond to languages supported by the model~\cite{Lietal2025Language}.
We classify comments into six categories using in-context learning with at least two examples per category: \emph{formatting} (navigation aids and visual separators), \emph{academic text} (commented-out paragraphs or alternative phrasings), \emph{conversational} (rather informal notes between coauthors), \emph{\LaTeX{} markup} (table structures or mathematical expressions), \emph{todo} (task lists and open items), and \emph{bibliography} (citation-related notes).
Additionally, we instruct the model to flag comments containing potentially private or sensitive information.
On a manually labeled test dataset of \num{500} random comments (stratified by LLM-predicted class), this method achieves a macro-averaged F1-score of \num{0.83} for assignment to categories, and \num{0.71} for flagging sensitive content.

\textbf{LLM-based Classification.}
Our evaluation shows that \emph{\LaTeX{} markup} comments are most prevalent (\qty{30}{\percent}, \qty{92}{\percent} of submissions, \impL), anecdotally containing table formatting, mathematical expressions, or disabled code.
The second most common category is \emph{academic text} (\qty{44}{\percent}, \qty{76}{\percent} of submissions, \impL), comprising commented-out paragraphs, alternative text formulations, or previous versions of arguments.
Manual inspection of selected cases reveals anecdotal evidence of conflicting lines of argumentation between published and commented text, though we leave a systematic analysis of semantic contradictions for future work.
\emph{Conversational} comments rank third (\qty{7.9}{\percent}, \qty{72}{\percent} of submissions, \impM) and frequently contain discussions among coauthors, including critical judgments about text quality, suggestions for improvement, or strategic considerations about presentation---content presumably not intended to be public.
Less common categories are \emph{formatting} aids (\qty{1.7}{\percent}, \qty{64}{\percent} of submissions, \impL), \emph{bibliography}-related comments (\qty{9.1}{\percent}, \qty{62}{\percent} of submissions, \impL), and \emph{todo} items (\qty{0.98}{\percent}, \qty{44}{\percent} of submissions, \impM).
For the latter category, we cannot automatically assess whether these tasks were ultimately addressed or not.
However, manual examination reveals cases where todos explicitly mention methodological limitations or acknowledge weaknesses that are not disclosed in the published text.

The model also flags \qty{7.0}{\percent} of comments as sensitive, including PII, opinions about reviewers/related work, author discussions, or credentials (\impM--\impH).
Via bootstrapping ($n=10 000$, \qty{95}{\percent} confidence) of the labeled dataset, we estimate that at least \qty{3.6}{\percent} indeed contain sensitive information.

Comparing the different comment mechanisms (\cf{} \cref{sec:methodology}), we find that most comments are standard line comments using \texttt{\%} (\qty{70}{\percent}).
We also identify comment blocks within \texttt{comment} environments (\qty{15}{\percent}), content in skipped branches of conditional statements (\qty{0.14}{\percent}), arguments discarded by custom commands (\qty{8.8}{\percent}), post-\texttt{document} content (\qty{0.085}{\percent}), and text within special environments (\qty{0.055}{\percent}).
Cross-referencing these mechanisms with our semantic categories reveals some notable patterns: \emph{conversational} comments appear disproportionately often in comment environments (\qty{15}{\percent}), whereas \qty{83}{\percent} of \emph{todo} comments are simple line comments, suggesting that this mechanism is primarily used for tracking tasks during collaborative writing.

\textbf{Disclaimer.}
These numbers should be taken with a grain of salt as
\begin{enumerate*}[label=(\alph*)]
	\item LLM classification can introduce misclassification, partly as we use a weaker self-hosted model for security reasons (\cf{} \cref{app:sec:llmusage}) and
	\item the perception of what is sensitive varies substantially across individuals~\cite{Leonetal2013What}.
\end{enumerate*}
Still, we believe that this approach is the best one available as manual annotation not only results in annotator bias but also fails to scale to millions of comments.

\textbf{Takeaway.}
\emph{%
Striking individual cases span a wide spectrum, including cryptographic material, credentials and PII of survey participants, editable online documents, (GPS) metadata from authors, unfinished tasks on todo lists, traceable submission histories, and collaboration artifacts.
Failed-to-delete NFS files and evidence of ineffective censorship emphasize the authors' intent to hide such information.
}

\section{Mitigation Through \LaTeX{} Sanitization Tools}
\label{sec:latexcleaner}

At first glance, the prevalence of ``hidden'' information might appear surprising, as different tools to sanitize (or ``clean'') \LaTeX{} sources---before submitting them to \arxiv{} or similar services---and thus mitigate the identified issues are readily available, some even actively advertised by \arxiv{}~\cite{arXiv2005Requiring}.

To establish a structured understanding of why these existing \LaTeX{} sanitization tools fail to address the issue of ``hidden'' information in \arxiv{} submissions, we first survey the landscape of cleaner tools (\cref{subsec:latexcleaner:comparison}) and then evaluate their performance in cleaning \LaTeX{} sources as well as estimate their usage frequency (\cref{subsec:latexcleaner:discovery}).
To address the inherent shortcomings of existing sanitization approaches our analysis identifies and to provide mitigations for the problem of ``hidden'' information uncovered by our work, we present \alcpp{}, our new sanitization tool (\cref{subsec:latexcleaner:mitigation}).

\subsection{Landscape of \LaTeX{} Sanitization Tools}
\label{subsec:latexcleaner:comparison}

\begin{table*}[t]
	\centering
	\caption{%
		Comparison of available tools for source file cleaning (sorted by technology and endorsement by \arxiv{}~\cite{arXiv2005Requiring}).
		None of these tools is able to reliably sanitize all test cases.
		Regardless, a subset of authors seems to apply them prior to submission.
	}
	\vspace{-1.0em}
	\begin{scriptsize}
	\begin{minipage}{0.8\textwidth}
	\setlength{\tabcolsep}{2.25pt}
	\begin{tabularx}{\textwidth}{
		>{\raggedright\arraybackslash}p{3.15cm}
		>{\raggedright\arraybackslash}p{2.25cm}
		>{\raggedright\arraybackslash}p{1.65cm}
		r
		| ccccccccc
		| rr
	}
		\toprule
		\textbf{} & \textbf{} & \textbf{Claimed} & \multicolumn{1}{c}{\textbf{Last}} & \multicolumn{9}{c}{\textbf{Comment Cleanup Tests}} & \multicolumn{2}{c}{\textbf{Sanitization}}\\
		\textbf{Name} & \textbf{Technology} & \textbf{Features} & \textbf{Update} & \ding{182} & \ding{183} & \ding{184} & \ding{185} & \ding{186} & \ding{187} & \ding{188} & \ding{189} & \ding{190} & \textbf{``Beneficial''} & \textbf{Breaks}\ensuremath{^\mathsection} \\
		\midrule

		perl one-liner~\cite{arXiv2005Requiring} &
			regex-based &
			comments &
			11/2005 & 
			\cmark{} & 
			\xmark{} & 
			\xmark{} & 
			\xmark{} & 
			\xmark{} & 
			\xmark{} & 
			\xmark{} & 
			\xmark{} & 
			\xmark{} & 
			\qty{95}{\percent} & 
			\qty{4.2}{\percent} \\ 

		arxiv\_latex\_cleaner (ALC)~\cite{Google-Research2019arxivlatexcleaner} &
			regex-based &
			\tabitem dangling files &
			03/2026 & 
			\cmark{} & 
			\xmark{} & 
			\cmark{} & 
			\xmark{} & 
			\cmark{} & 
			\cmark{} & 
			\xmark{} & 
			\manualmark{} & 
			\cmark{} & 
			\qty{91}{\percent} & 
			\qty{19}{\percent} \\ 
		 & 
			& 
			\tabitem comments &
			& 
			\multicolumn{9}{r|}{\tiny\emph{with manually-enabled command cleaning:}} &
			\qty{98}{\percent} & 
			\qty{28}{\percent} \\ 

		latexindent.pl~\cite{Hughes2012latexindent.pl} &
			regex-based &
			comments &
			02/2026 & 
			\cmark{} & 
			\cmark{} & 
			\xmark{} & 
			\xmark{} & 
			\xmark{} & 
			\xmark{} & 
			\xmark{} & 
			\xmark{} & 
			\xmark{} & 
			\qty{3.7}{\percent} & 
			\qty{9.2}{\percent} \\ 

		\arxiv{} Cleaner~\cite{Elsa-Lab2019arXiv} & 
			\tabitem pdflatex recorder &
			\tabitem dangling files &
			09/2019 & 
			\cmark{} & 
			\cmark{} & 
			\crashmark{} & 
			\xmark{} & 
			\cmark{} & 
			\xmark{} & 
			\xmark{} & 
			\xmark{} & 
			\cmark{} & 
			\qty{80}{\percent} & 
			\qty{80}{\percent} \\ 
		 & 
			\tabitem latexpand~\cite{Moy2012latexpand} &
			\tabitem comments &
			& 
			\multicolumn{9}{r|}{\tiny\emph{only latexpand:}} & 
			\qty{69}{\percent} & 
			\qty{85}{\percent} \\ 

		Sub.\ Sanitizer \& Flattener~\cite{Stutz2022Submission} &
			snapshot~\cite{American-Mathematical-Society1999snapshot} &
			dangling files &
			06/2022 & 
			\xmark{} & 
			\xmark{} & 
			\xmark{} & 
			\xmark{} & 
			\xmark{} & 
			\xmark{} & 
			\xmark{} & 
			\xmark{} & 
			\xmark{} & 
			\multicolumn{2}{c}{\tiny\emph{not reasonably applicable}}\\

		pandoc~\cite{MacFarlane2006Pandoc} &
			tokenization &
			n/a &
			03/2026 & 
			\cmark{} & 
			\cmark{} & 
			\cmark{} & 
			\cmark{} & 
			\cmark{} & 
			\xmark{} & 
			\cmark{} & 
			\cmark{} & 
			\xmark{} & 
			\multicolumn{2}{c}{\tiny\emph{not reasonably applicable}}\\

		\midrule

		\alcpp{}~\cite{Pennekampetal2026ALC-NG} \emph{\tiny(this work)} &
			\tabitem pdflatex recorder &
			\tabitem dangling files &
			04/2026 & 
			\cmark{} & 
			\cmark{} & 
			\cmark{} & 
			\cmark{} & 
			\cmark{} & 
			\cmark{} & 
			\cmark{} & 
			\cmark{} & 
			\cmark{} & 
			\qty{98}{\percent} &
			\qty{15}{\percent} \\
		& 
			\tabitem exiftool~\cite{Harvey2003ExifTool} &
			\tabitem metadata &
			& 
			\multicolumn{9}{r|}{} &
			\multicolumn{2}{r}{} \\
		& 
			\tabitem tree-sitter-based~\cite{tree-sitter2018Tree-sitter} &
			\tabitem comments &
			& 
			\multicolumn{9}{r|}{} &
			\multicolumn{2}{r}{} \\

		\bottomrule
	\end{tabularx}
	\centering
	\;\\[-0.25em]
	\cmark{} cleans \LaTeX{} source files successfully, \xmark{} unsuccessful (attempt), \manualmark{} requires manual action for sanitization, and \crashmark{} tool crashes, \ie{} failed test
	\end{minipage}
	\hfill
	\begin{minipage}{0.19\textwidth}
	\;\\[0.25em]
	\underline{\emph{Comment Cleanup Tests:}}
	\begin{dingautolist}{182}
	\setlength{\leftmargin}{0em}
	\setlength{\itemindent}{-1.5em}
	\setlength{\labelwidth}{1.5em}
	\setlength{\labelsep}{0.5em}
		\item Inline comment removal
		\item \textbackslash\textbackslash\% detection
		\item Comment environment
		\item Retain comments within\newline special environments
		\item Out of document removal
		\item \textbackslash iffalse/\textbackslash if0 handling
		\item Custom \textbackslash if removal
		\item No arg.\ command cleaning
		\item Bbl file support
	\end{dingautolist}
	\pgfkeys{/csteps/inner ysep=2pt} \pgfkeys{/csteps/inner xsep=2pt}
	Mapped to \cref{sec:methodology} (concepts):\newline
	\nostep{A}:\,\ding{182}--\ding{185};\,\nostep{B}:\,\ding{186};\,\nostep{C}:\,\ding{187}--\ding{188};\,\nostep{D}:\,\ding{189}\\[0.35em]
	{\tiny
		\ensuremath{^\mathsection}(i) tool crashes, (ii) paper cannot compile, or\\[-0.45em]
		\hspace*{2em}(iii) mismatch in pixel-perfect comparison
	}
	\end{minipage}
	\end{scriptsize}
	\vspace{-1.25em}
	\label{tab:source-cleaner}
\end{table*}

Different so-called \emph{sanitization tools} promise to ``clean'' \LaTeX{} source files to prevent issues stemming from ``hidden'' information when preparing submissions, \eg{} to \arxiv{}.
In the following, we provide a structured and comprehensive overview of the various sanitization approaches and provide a first assessment of their cleaning capabilities.

To identify tools and software designed for cleaning \arxiv{} source files, we systematically searched for projects on GitHub using the keywords \texttt{\arxiv{} cleaner} OR \texttt{arXiv sanitizer}.
We further considered a compilation of different approaches on StackExchange from 2012~\cite{StackExchange2012Utility}.
We then selected representative approaches, one per underlying technology, based on popularity (stars on GitHub) and compiled a detailed comparison of these tools in \cref{tab:source-cleaner}.

Out of these sanitization tools, the following warrant further discussion due to their popularity or unique approach.

\emph{arxiv\_latex\_cleaner (ALC)}~\cite{Google-Research2019arxivlatexcleaner}.
Arguably, this cleaner developed by Google is the most prominent one, operating both on the \emph{file level} and the \emph{content/text level}, as evidenced by various recommendations~\cite{wwwwi112023} and \arxiv{}'s endorsement since 12/2019~\cite{arXiv2005Requiring}.
Different academic works even apply it to preprocess input data as part of their research methodology in the context of LLM/AI research~\cite{Maatouketal2024Tele-LLMs:,Abramovichetal2025AblationBench:,Lietal2025WirelessMathBench:}.
Despite ongoing efforts to ensure correct behavior~\cite{sbirchfield2020Should}, several open issues related to the accurate detection (and removal) of comments, \ie{} irrelevant content, remain~\cite{refraction-ray2019Feature,albertfgu2020Deletes}.

\emph{Submission Sanitizer \& Flattener}~\cite{Stutz2022Submission}.
In contrast, the Submission Sanitizer \& Flattener focuses specifically on the \emph{file level}, \ie{} solely removing files that are not needed, but not considering the irrelevant content and text contained in the remaining files.
As such, it differs quite substantially from the other tools in our comparison.
To identify which files are needed to build a \LaTeX{} document, an additional build process is required using the \texttt{snapshot} package~\cite{American-Mathematical-Society1999snapshot}.

\emph{pandoc}~\cite{MacFarlane2006Pandoc}.
While not advertised as a sanitization tool, pandoc, as a universal document converter, can also clean \LaTeX{} sources at the \emph{content/text level}.
However, it massively reformats the sources, adds new \LaTeX{} instructions, and breaks compatibility with common templates (\eg{} \texttt{acmart}), severely challenging its usage, let alone its automated usage.

Given this diverse landscape of sanitization tools, we sought to better understand their capabilities.
To this end, using synthetic experiments, we assess their ability to remove irrelevant content alongside the four concepts for embedding comments in \LaTeX{} documents we derived in \cref{sec:methodology} (\nostep{A} comments, \nostep{B} outside \texttt{document}, \nostep{C} non-taken \texttt{if} branches, \nostep{D} unused arguments of commands).
We created eight distinct test cases (\ding{182}--\ding{189}) that comprehensively cover these four concepts (\cf{} \cref{tab:source-cleaner}), including straightforward and more subtle ways of irrelevant content as well as content that might be falsely detected as a comment (\eg{} text within \texttt{verbatim} environments).
Additionally, we test whether sanitization tools replace calls to \BibTeX{} with the content of the corresponding \texttt{bbl} file (\ding{190}).
Our research artifact~\cite{Pennekampetal2026Artifact:} contains a minimal working example that covers these cases.

The results of our synthetic comment cleanup tests listed in \cref{tab:source-cleaner} show that none of the sanitization tools perform satisfactorily in terms of cleaning coverage, despite \arxiv{} endorsing two of them~\cite{arXiv2005Requiring}.
Even among tools that utilize the same technology, \eg{} regex-based ones, we observe notable differences in their cleanup coverage.
Surprisingly, the \arxiv{} Cleaner~\cite{Elsa-Lab2019arXiv} even crashes when encountering a basic \texttt{comment} environment in the source file.
Additionally, to remove certain irrelevant content, some sanitization tools require manual interaction, which impairs usability for inexperienced \LaTeX{} users.
Moreover, all cleaners but ALC~\cite{Google-Research2019arxivlatexcleaner} require the user to explicitly identify the primary \LaTeX{} file, potentially constraining the ease of use.
This situation is somewhat unexpected since \arxiv{}'s submission system can process submissions automatically, even if it means falling back to the \arxiv{}-specific \texttt{00README}(\texttt{.XXX}) file~\cite{arXiv2010The}.

All available cleaners exclusively cover the \emph{file level} and (certain aspects of) the \emph{content/text level}, ignoring (sensitive) content at the \emph{metadata level}.
Hence, they entirely disregard a complete dimension (\cf{} \cref{sec:methodology}) to the problem of ``hidden'' information in \LaTeX{} sources.

\subsection{Quantifying Sanitization Performance}
\label{subsec:latexcleaner:discovery}

While our analysis of \LaTeX{} sanitization tools (\cf{} \cref{subsec:latexcleaner:comparison}) provides a \emph{theoretical} understanding of their capabilities (and limitations), a \emph{practical} quantification of their sanitization performance on actual \arxiv{} submissions is still missing.
Thus, we now evaluate
\begin{enumerate*}[label=(\alph*)]
	\item whether the application of sanitization tools would have removed additional ``hidden'' information in \arxiv{} submissions,
	\item which ``hidden'' information still remains after sanitization,
	\item whether sanitization tools have been applied on sources prior to submission, and
	\item how large the potential for storage savings at \arxiv{} is if sanitization tools were applied consistently.
\end{enumerate*}

\textbf{Methodology.}
To answer these questions, we run the identified cleaners (\cref{tab:source-cleaner}) on our dataset (\cref{sec:dataset}).
To establish a reliable baseline for dangling files, we monitor file system accesses during the build process and classify files that have not been accessed as dangling.
For the build process, we utilize the most recent \TeX{} Live version in Ubuntu 24.04 (\num{2023}) and latexmk~\cite{Collins1998latexmk} (v4.83).
Due to breaking changes over time~\cite{Tanetal2024Inconsistencies}, both in \TeX{} Live and author-included \LaTeX{} packages, we are unable to successfully compile \emph{all} \LaTeX{} sources.
Still, we can compile---and thus evaluate the different sanitization tools on---\qty{2.32}{\million} papers (\qty{85}{\percent}).

\textbf{Sanitization Performance.}
To quantify whether the application of sanitization tools would have been beneficial, \ie{} removes additional ``hidden'' information, we apply the different cleaners and evaluate whether they reduce the amount of dangling files and comments for a given paper.
\Cref{tab:source-cleaner} reports
\begin{enumerate*}[label=(\alph*)]
	\item the share of papers for which a sanitization tool alters the sources, with the paper still compiling after application (\emph{``beneficial''}---theoretical benefit, albeit no guarantee of full sanitization coverage or accurately retaining visible content), and
	\item the share of papers where the tool \emph{breaks} the sanitization attempt, either via tool failure, removal of essential content, or other issues that visually alter the document.
\end{enumerate*}
Exemplarily focusing on ALC as the most prominent tool, \qty{91}{\percent} of the evaluated submissions (\qty{2.12}{\million}) would ``benefit'' in terms of (additionally) removed ``hidden'' information.
In detail:
For \qty{50}{\percent}, the number of dangling files decreases; similarly, for \qty{91}{\percent}, the number of comments decreases.
ALC supports manually specifying commands that should be removed.
Even though we expect laypersons to rarely make use of this manual feature, we report numbers for both behaviors in \cref{tab:source-cleaner} as well.
Importantly, these results do not entail that ALC perfectly sanitizes those papers.
Rather, it merely \emph{reduces} the amount of ``hidden'' information.
Sometimes, it also \emph{overremoves relevant} content, leading to an inflated sense of beneficialness, \ie{} ``breaking'' the content.
Similar observations hold for the other sanitization tools.

Especially when identifying (and removing) dangling files, relying on regexes (as done by ALC) comes with the inherent risks of a heuristic approach, \ie{} potentially leading to over- or underremoval.
As a result, \LaTeX{} sources do not compile anymore, or dangling files---other than intended---are not removed.
Quantifying this issue, we also visualize the difference between using the file system access-based approach (\cf{} methodology) as a reliable baseline and ALC in \cref{fig:cleaner-performance} (\cref{app:sec:evalextra}).
Similar issues arise for regex-based comment removal, where special environments, such as \texttt{listing} or \texttt{verbatim}, are frequently mismanaged, leading to ``clean'' sources with altered content.

Consequently, while existing tools are beneficial in that they remove sensitive information, they also come with the risk of breaking \LaTeX{} sources and altering content, albeit not comprehensively removing sensitive information.

\textbf{Estimating Attempted Sanitization.}
To roughly quantify this false sense of security, we check whether applying a sanitization tool's most recent version results in unchanged sources, which may indicate previous use.
We find \qty{106}{\thousand} submissions without any comments and/or dangling files, which authors might have sanitized (manually or using a sanitization tool).
Still, the large number of papers where cleaners are considered ``beneficial'' highlights that only few authors were able to successfully apply them.
Trivial line comments, which every sanitizer passing \ding{182} removes accurately (\cf{} \cref{tab:source-cleaner}), in \qty{94}{\percent} of evaluated submissions suggest that no comprehensive attempt was made.
Considering the failure to handle \ding{183} as a fingerprint of ineffective ALC usage, we find \num{538} uses (\qty{0.10}{\percent} of submissions since its release in 2018~\cite{Google-Research2018History}), \ie{} a minuscule prevalence of confirmable sanitization with an \arxiv{}-endorsed tool~\cite{arXiv2005Requiring}.

\textbf{Remaining Sensitive Information.}
We now focus on any remaining sensitive information \emph{after} applying sanitization tools.
In addition to the underremoval of dangling files, we identify thousands of remaining, \ie{} underremoved, \emph{comments} after running ALC.
Within those comments, our regex-based detection (\cf{} \cref{subsubsec:analysis:individual:regexable}) reveals \num{362} matches, \eg{} including \num{7} (sensitive) links to Overleaf, highlighting the need for ``perfect'' removal.
Consequently, current sanitization tools might create a false sense of security by potentially \emph{not} removing all occurrences of sensitive information.

\textbf{Storage Overhead.}
Besides unintentionally disclosing ``hidden'' information, missing out on potential for sanitizing unnecessary information from \arxiv{} submissions also puts a \emph{permanent} storage burden on \arxiv{}.
For the set of \qty{2.32}{\million} submissions we can compile, we identify \qty{658}{\giga\byte} (\qty{22.9}{\percent}) that could be freed by accurate sanitization of the analyzed version only; \qty{97}{\percent} account for dangling files, \qty{0.4}{\percent} for metadata, and \qty{2.3}{\percent} for content in required files.

\subsection{\alcpp{} -- A Reliable \LaTeX{} Sanitization Tool}
\label{subsec:latexcleaner:mitigation}

Motivated by the shortcomings of current \LaTeX{} sanitization tools and following our ethical obligation to provide a reliable mitigation for the threat posed by ``hidden'' information in \LaTeX{} sources, we open-source our ``\textbf{N}ext-\textbf{G}eneration'' sanitization tool, called \alcpp{}~\cite{Pennekampetal2026ALC-NG} (See: \underline{\url{http://alc-ng.de}}).

With this tool, we combine established best practices for each dimension of ``hidden'' information (\cf{} \cref{sec:methodology}), \ie{} pdflatex recorder for dangling file detection, exiftool for metadata removal, and a tree-sitter-based detection of irrelevant content (comments).
While \arxiv{} Cleaner~\cite{Elsa-Lab2019arXiv} also applies pdflatex recorder, state-of-the-art sanitization tools do not yet use exiftool or a tree-sitter grammar (\cf{} \cref{tab:source-cleaner}).

\nostep{1}{} \emph{File Level.}
To reliably identify files that are required to compile a \LaTeX{} document, we utilize \emph{pdflatex recorder} (similar to \arxiv{} Cleaner~\cite{Elsa-Lab2019arXiv}).
Using the access timestamp (\emph{atime}) of the file's last read~\cite{Free-Software-Foundation2017File}, we validated the correctness of this approach on the \qty{2.32}{\million} compilable papers by comparing its output with read accesses on disk and did not discover any discrepancies.
Furthermore, we use the \texttt{00README} file~\cite{arXiv2010The} as well as the directory \texttt{/anc/} for ancillary files~\cite{arXiv2010Ancillary} to identify files \emph{not} to remove.

\nostep{2}{} \emph{Metadata Level.}
To sanitize metadata, we utilize \emph{exiftool}~\cite{Harvey2003ExifTool} since it is known for reliably removing metadata across various file types, including PDFs and images.
Optionally, we support suppressing file modification times to hide corresponding sensitive behavioral metadata further.

\nostep{3}{} \emph{Content/Text Level.}
To reliably identify and remove unnecessary content, we re-purpose our abstract syntax tree-based detection logic (\cf{} \cref{sec:methodology}).
As we summarize in \cref{tab:source-cleaner}, this approach comprehensively covers all aspects of our synthetic comment cleanup tests.

To validate the correctness of \alcpp{}'s sanitization, \ie{} ensure that the removal of files, metadata, and content does not change the resulting compiled PDF, we compare the original and the sanitized PDFs using \emph{pdfium}~\cite{pdfium2014PDFium} to ensure \emph{pixel-perfect} versions, \ie{} visually-identical PDFs.
We get visually-identical PDFs for \qty{87}{\percent} of the compilable submissions \emph{after sanitization}; as opposed to \qty{96}{\percent} for the perl one-liner as the best cleaner for this metric, but subject to underremoval (little sanitization coverage), and \qty{88}{\percent} for ALC, albeit for a lower number of submissions that remain compilable \emph{after sanitization}.
Applying the current prototype of \alcpp{} on the \qty{2.32}{\million} compilable submissions shows that \qty{98}{\percent} (\qty{2.27}{\million}) benefit from its use by excluding dangling files and/or cleaning irrelevant content:
\alcpp{} comprehensively removes the different dimensions of ``hidden'' information within \LaTeX{} sources (\cf{} \cref{sec:methodology}).
Despite its proof-of-concept implementation, it outperforms the surveyed sanitization tools in terms of coverage (\cf{} \Cref{tab:source-cleaner}), while promising great ease of use even for inexperienced \LaTeX{} users, laying the foundation for broad adoption.

Consequently, with \alcpp{}, we empower authors to reliably sanitize their \LaTeX{} sources before submitting them to \arxiv{} (or other third parties) and thus mitigate the various threats posed by ``hidden'' information identified in this paper.

\section{Discussion and Implications}
\label{sec:discussion}

After concluding our analysis of ``hidden'' information and sanitization tools, we now focus on permeating our findings.
Specifically, we discuss key implications (\cref{subsec:discussion:discussion}), perception in the community based on a survey among affected \arxiv{} authors (\cref{subsec:discussion:userstudy}), limitations of our work (\cref{subsec:discussion:futurework}), \arxiv{}-specific aspects (\cref{subsec:discussion:arxiv}), and recommendations for stakeholders (\cref{subsec:discussion:recommendations}).

\subsection{Putting Our Findings into Perspective}
\label{subsec:discussion:discussion}

Throughout our work, we considered the guiding questions:
\emph{%
How prevalent are the three dimensions of ``hidden'' information in source files of preprints?
Which individual cases of ``hidden'' information can we discover?
How sensitive are these cases?}
We now revisit these questions.

\textbf{General Assessment.}
Our analysis uncovered a wide range of ``hidden'' information.
We discovered information with immediate confidentiality and security implications, as well as details that can be used to deduce behavioral aspects, \eg{} timestamps or sanitization attempts.
Hence, the issues we identify vary greatly in severity.
Even seemingly trivial attempts to mitigate the issue---accurately removing dangling files prior to submission---would have removed \qty{20}{\percent} of links, \qty{33}{\percent} of GPS locations, and \qty{5.9}{\percent} of unique comments.
On another note, perhaps counterintuitively, papers appearing on both \arxiv{} and in A\textsuperscript{*}- and A-ranked security conferences disclose, on average, more ``interesting'' and sensitive information than their counterparts from other areas, despite security researchers' presumed awareness of such risks.

In \cref{app:sec:evalextra}, we further complement this paper by discussing ``hidden'' information in our \emph{own} \arxiv{} papers.
Moreover, we revisit the \mbox{GPT-4} paper (\cf{} \cref{sec:introduction}) to assess whether that case could have been easily prevented.

\textbf{Risk of Breaking Blindness in Peer Review Processes.}
Our findings have implications beyond preprint servers, potentially affecting (double-blind) review and publication processes where authors supply source files.
Particularly, beyond the well-known implications of metadata, reviewers or infrastructure providers could exploit additional ``hidden'' information once more reviewing and publication processes require the submission of \LaTeX{} source files.
Relatedly, the rise of collaborative \LaTeX{} editors~\cite{Lacombeetal2021Can} will further increase the usage and distribution of \LaTeX{} sources.

\subsection{Responsible Disclosure and Author Survey}
\label{subsec:discussion:userstudy}

Throughout our research, we responsibly disclosed validated security-critical findings to \num{2660} authors of affected papers, informing them about our findings and allowing them to take corrective actions (\cf{} \cref{app:sec:ethics} for more details on our considerations surrounding the responsible disclosure).

We complemented our disclosure notifications with an overview of other sensitive findings (\eg{} potentially sensitive \LaTeX{} comments) and a link to a survey to assess awareness and practices regarding \LaTeX{} sources when submitting to \arxiv{}.
We received \num{112} complete survey responses (response rate: \qty{5.3}{\percent}).
The largest group of respondents were assistant (\qty{29}{\percent}) or full (\qty{16}{\percent}) professors, followed by permanent research staff (\qty{28}{\percent}) and postdoc/PhD students (\qty{18}{\percent}).
\qty{49}{\percent} of respondents claimed more than ten \arxiv{} submissions, and only \qty{41}{\percent} stated awareness of source file availability and its implications before receiving our disclosure notice.

\textbf{Sensitivity.}
\qty{43}{\percent} of respondents agreed that the disclosed information is sensitive.
However, when asking specifically about the dimensions of information leakage, only \qty{21}{\percent} confirmed that no sensitive information was contained in their sources, indicating disagreement and uncertainty about what constitutes sensitive information (\cf{} disclaimer in \cref{subsubsec:analysis:individual:comments}).
For the other respondents, \emph{conversational} comments were considered sensitive by \qty{39}{\percent}, and \emph{todo} comments by \qty{38}{\percent} of respondents (\cf{} \cref{subsubsec:analysis:individual:comments}).
For the \qty{21}{\percent} reporting no sensitive information, we find significantly fewer dangling files (-\qty{74}{\percent}), metadata values (-\qty{40}{\percent}), and unique comments (-\qty{36}{\percent}, Mann-Whitney-U test: $p\leq 0.03$ in all cases).
This finding suggests that authors possess some intuition regarding sensitive hidden information.
\num{9} email respondents stated that disclosed online documents did not contain sensitive information, but \num{7} still restricted access.
By March 2026, we discovered \num{18} updated submissions that followed our disclosure, indicating cleanup attempts.
Irrespectively, the affected, old source files remain accessible.

\textbf{Perceived Helpfulness.}
\qty{57}{\percent} of respondents indicated that the disclosure was very helpful (\qty{79}{\percent} when including participants who found it rather helpful).
Separate from survey participation, \num{51} authors replied positively to our disclosure via email.
In contrast, two respondents reported false positives from our detection, requesting that we should have invested manual effort to verify the issue---which we thoroughly did (\cf{} \cref{subsubsec:analysis:individual:regexable}).

\textbf{Experience with Cleaners.}
Among respondents, \qty{28}{\percent} had heard of cleaning tools listed in \cref{tab:source-cleaner}, \qty{12}{\percent} had tried them, and \qty{11}{\percent} report using at least one cleaner regularly.
Regarding cleaning practices, \qty{8.9}{\percent} reported applying (manual) cleaning before the submission, which we quantify as reducing \qty{36}{\percent} of unique comments in the notified-upon submissions; however, \emph{all} of these submissions still had issues.
Seven authors responded that, following our disclosure, they uploaded a new, cleaned version---not realizing that old versions and sources remain accessible.

\textbf{Potential Biases.}
We acknowledge potential bias in our survey responses, since we primed participants immediately before participation by indicating that we considered their sources to leak potentially sensitive information.
Moreover, we only contacted authors who disclose publicly-accessible Google Docs URLs or credentials (manual curation) within irrelevant content, disregarding (more careful) authors without such findings.
Likewise, authors who did not consider our findings sensitive might have been less inclined to respond to our survey and/or did not expect other sensitive findings.
Still, we received text responses containing more details than disclosed in our email, indicating that at least some authors carefully reflected on our disclosure campaign.

\subsection{Limitations and Implications}
\label{subsec:discussion:futurework}

While our methodology enables the study of ``hidden'' information in \arxiv{} sources, it is not without limitations.

These limitations mainly stem from the scale of the research problem, which demands different assessment directions as evident in the diversity of our analysis (\cf{} \cref{subsec:analysis:individual}).
First, we pursue a best-effort approach when analyzing the sources and compiling paper PDFs, which comes with some minor limitations:
\begin{enumerate*}[label=(\roman*)]
	\item \num{1} source file archive is corrupt.
	\item We limit both parsing and compiling of source files to \qty{5}{\minute} and kill the compiler once its logfile exceeds \qty{2}{\giga\byte}, resulting in \num{388} prematurely terminated analyses.
	\item We ignore parsing errors when processing the sources using the tree-sitter grammar (\cf{} \cref{sec:methodology}), affecting \qty{681}{\thousand} submissions, resulting in, on average, \num{9829} bytes of disregarded content that may contain ``hidden'' information.
\end{enumerate*}

Second, our comment detection within \LaTeX{} files revealed \qty{644}{\million} occurrences.
While ideally we would have taken a manual look at \emph{all} individual comments, this approach is clearly infeasible.
Instead, we combined our definition of ``interesting'' content, various regex detection methods, and an LLM-based analysis of author comments to gain a comprehensive picture.
Arguably, this reasonable best-effort approach may have missed important sensitive details and other issues.
As such, the details presented in this paper ``only'' provide an accurate lower bound concerning the prevalence of ``hidden'' information, potentially missing more findings.

For the implications of applying an LLM-based content classification, we refer to our disclaimer in \cref{subsubsec:analysis:individual:comments} and our LLM usage considerations in \cref{app:sec:llmusage}.

Likewise, as our assessment could only focus on source files that remain publicly accessible, we could not consider submissions and versions where sources have purposefully been removed.
Thus, our version-related conclusion (\cf{} \cref{subsubsec:analysis:bigpicture:differences}) may be skewed, since very sensitive source files may have actively been removed.
Only if \arxiv{} persisted those sources, they could still conduct such an analysis internally.
Irrespectively, third-party researchers cannot address this data collection-inherent limitation for very good reasons.

Finally, for our evaluation of existing sanitization tools (\cf{} \cref{subsec:latexcleaner:comparison}), we constrained our analysis to \arxiv{} submissions that we could compile in a modern \LaTeX{} build environment (\cf{} \cref{subsec:latexcleaner:discovery}).
As a result, our reported numbers cover \qty{85}{\percent} of all submissions with sources.
For the remainder, evolving \LaTeX{} compilers and breaking changes in packages prevent compilation~\cite{Tanetal2024Inconsistencies}.
Still, by covering the majority of submissions with sources in our analysis, we are confident that it accurately reflects the performance of the different sanitization tools, especially considering that these tools will (only) be applied to new \arxiv{} submissions and thus will compile within a modern \LaTeX{} build environment.

\subsection{Identified \arxiv{}-specific Peculiarities}
\label{subsec:discussion:arxiv}

Given our focus on \arxiv{}, we also noted and considered some of its specific peculiarities, as detailed below.

\begin{figure}[t]
	\centering
	\includegraphics[width=\linewidth,trim={0 0 0 0.25em},clip]{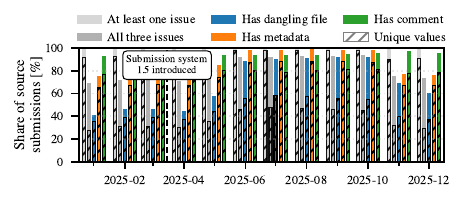}
	\vspace{-2.5em}
	\caption{%
		The share of submissions that feature ``hidden'' information is consistent over time in 2025.
		Additionally, we do not see any notable impact following the introduction of \arxiv{}'s new submission system.
	}
	\vspace{-0.75em}
	\label{fig:currentyear}
\end{figure}

\textbf{Implications of the new Submission System.}
In April 2025, \arxiv{} introduced a new submission system (1.5).
We can easily track the usage of this submission system through the values of the \texttt{Producer} metadata key.
Even though it introduces notable changes to submission processing, we do not observe any immediate implications on our findings:
\Eg{} the new system now specifically prompts submitters to remove dangling files on the root-level, while other directories and their dangling files remain unaffected.
\cref{fig:currentyear} (\cref{app:sec:evalextra}) shows a zoomed-in version of our analysis of ``hidden'' information for 2025, indicating no substantial changes through the switch to the new submission system.

\textbf{\arxiv{}-specific Cleaner Functionality.}
\alcpp{}---different from previous sanitization tools ``for'' \arxiv{}---supports \arxiv{}-unique functionality.
Thus, we support both \texttt{00README}(\texttt{.XXX}) files~\cite{arXiv2010The} (used to override defaults) and \texttt{anc/} directories~\cite{arXiv2010Ancillary} (to intentionally add ancillary files).

\textbf{Identifying Exploitation of Findings.}
Complementing our assessment of \arxiv{} sources, we were interested in identifying whether other parties investigate such ``hidden'' information as well.
To this end, we added ``honey pot'' URLs to an otherwise benign \arxiv{} submission within the CS category.
More specifically, we added multiple URLs to the paper across all three dimensions of ``hidden'' information (\cf{} \cref{sec:methodology}) and monitored HTTP(S) requests.
However, since its submission to \arxiv{} in 05/2025, we have only recorded accesses to URLs that are embedded within the compiled PDF file.
URLs that are solely part of the sources have not been accessed yet.
While limited in scope, this observation suggests that, since deployment, no large-scale analysis with a similar methodology as ours (involving accessing discovered URLs) processed our ``honey pot'' submission in detail.

\textbf{Meta Aspects.}
Given that the majority of \arxiv{} submissions features ``hidden'' information, partially leaking very sensitive details, future work should focus on better educating and thus supporting authors.
This aspect is particularly important since \arxiv{} is currently not considering additional changes related to the (retrospective) sanitization of ``hidden'' information.
Once a tool like \alcpp{} has matured sufficiently, \arxiv{} might consider integration into its submission system.
Beyond the reliable removal of ``hidden'' information, a substantial reduction of storage needs (\cf{} \cref{subsec:latexcleaner:discovery}) could be another driver for such a change.

\subsection{Recommendations for Stakeholders}
\label{subsec:discussion:recommendations}

Based on our findings and considering the prevalence and severity of ``hidden'' information in source files of preprints, we derive recommendations for affected stakeholders.

\textbf{Authors and Institutions.}
To ensure confidentiality of sensitive information and intellectual property, authors must ensure that they remove all comments and dangling files from their source files prior to submission to a preprint repository, preferably using a thorough sanitization tool such as \alcpp{}~\cite{Pennekampetal2026ALC-NG}.
Although not in the main focus of our work, this sanitization should include removing sensitive metadata from included PDF files and images (\cf{} \cref{subsubsec:analysis:individual:metadata}), using either dedicated tools~\cite{Harvey2003ExifTool,Berkenbilt2005qpdf,Voisin2018mat2} or (manually) orchestrated through \alcpp{}.
From an institutional perspective, to protect affiliated authors and intellectual properties, institutions should offer guidelines and establish precise policies on the release of \LaTeX{} sources.
Already affected authors and institutions, \ie{} those ones that (unknowingly) released not (properly) sanitized source files in the past, are likely unable to fully remove unintentionally-disclosed information (\cf{} \cref{app:sec:ethics}).
Consequently, corrective actions should concentrate on revoking permissions for, \eg{} exposed API keys and enforcing access control for affected URLs.

\textbf{Preprint Repositories.}
Complementing efforts by authors and institutions, preprint repositories should raise awareness for the risks resulting from ``hidden'' information and enforce sensible behavior.
For example, by incorporating checks on irrelevant content that sanitization tools would remove, preprint repositories could issue warnings for detected potentially sensitive ``hidden'' information in uploaded source files and automatically discard dangling files.
For already published source files, preprint repositories could proactively check for indicators of sensitive ``hidden'' information and work with authors to expurgate affected submissions.

To ease the understanding, we provide an executive summary of our findings at \underline{\url{https://arxiv.comsys.rwth-aachen.de}}.

\section{Conclusion}
\label{sec:conclusion}

In this work, we systematically assessed the issue of unintentionally-disclosed information in source files of \qty{2.7}{\million} \arxiv{} submissions.
Our findings reveal ``hidden'' information in almost every submission, albeit with greatly varying sensitivity.
Apart from unused template files that put unnecessary storage burden on \arxiv{}, we further discover scripts, research data, and even entire Git repositories.
Additionally, comments in \LaTeX{} sources reveal, \eg{} author conversations or todo items---for some of those comments, we are certain that the authors did not intend to disclose them publicly.
Alarmingly, our findings also include URLs without any access restrictions to other resources (\eg{} Google Docs), security tokens, and private keys.
Ultimately, the issue concerns three dimensions: dangling files, embedded metadata, and irrelevant content, \eg{} in comments.

Considering possible mitigations, our work reveals that current \LaTeX{} sanitization tools leave room for improving coverage and accuracy (resulting in both under- and overremoval).
Combining technologies known to perform well with a previously unconsidered approach to detect \LaTeX{} comments, we develop \alcpp{}, a reliable cleaner for source files.
Our findings further show that better education among researchers is needed to holistically address the issue of unintentionally-disclosed information within source files.

\section*{Acknowledgment}
\label{sec:acks}
Thank you to \arxiv{} for use of its open access interoperability and for supporting our research activities.
The authors thank Cloudflare (domain classification) and NVIDIA (compute activities) for supporting this work, and Muzamil Bashir for his initial, prototypical validation of this research.
We also thank the anonymous reviewers and our shepherd for their valuable feedback and suggestions.

\bibliographystyle{IEEEtran}
\bibliography{paper}

\appendices

\makeatletter
\renewcommand{\section}[1]{
  \refstepcounter{section}
  \phantomsection
  \vspace{1.0\baselineskip}
  {\noindent\normalfont\large\bfseries
   Appendix~\Alph{section}: #1\par}
  \vspace{0.5\baselineskip}
  \addcontentsline{toc}{section}{Appendix~\Alph{section}: #1}
  \markboth{Appendix~\Alph{section}: #1}{}
}
\makeatother

\crefalias{section}{appendix}

\section{Ethics Considerations}
\label{app:sec:ethics}

At its core, this research includes handling sensitive (confidential), potentially personally identifiable information, reviewer comments, credentials/key material, links/references to other resources, and software/scripts that are not meant for distribution, requiring thorough ethical considerations.

\textbf{Potential Harm.}
Most importantly, while we are only examining data that is purposely made publicly available, we still carefully align our research to maximize benefits while minimizing any potential harm.
By providing a substantially improved mitigation approach and responsibly disclosing any potentially serious findings, we minimize the potential negative impact of our research on a small number of individuals while at the same time providing a thorough understanding of and awareness of the risks involved with ``hidden'' information and thus benefiting all authors (submitting to \arxiv{}).
Unfortunately, given the available source file mirrors, including \url{archive.org} and the official arxiv-bulk dataset (\cf{} \cref{sec:dataset}), permanently restricting access to once published sources, even if authors manually request \arxiv{} to take them down---which alarmingly is an undocumented option---is challenging.
Thus, authors have to take other measures to deal with unintentionally-published, \ie{} compromised, information.
Accordingly, when publishing our analysis results~\cite{Pennekampetal2026Artifact:} as part of our commitment to open science, we obfuscate paper identifiers to prevent linkage of potentially sensitive data to individual submissions and their authors.
To further mitigate any negative impact of our work, we carefully ensured that it did not disrupt the operation of \arxiv{}.
In particular, we crawled any directly-retrieved data in compliance with \arxiv{}'s communicated request rates~\cite{arXiv2019Terms}.

\textbf{Accessing Web References.}
As part of our analysis in Section \ref{subsubsec:analysis:individual:regexable}, we manually accessed ``hidden'' URLs found in source files for two main reasons:
First, the mere presence of an URL neither reveals whether this URL is (still) accessible nor whether the URL points to sensitive information.
Second, by (manually) verifying whether referenced content is indeed sensitive and by disclosing such findings, we enable affected authors to correct their access control permissions, ideally preventing (continued) exploitation by malicious actors.

Unfortunately, accessing these URLs could expose us to sensitive or illegal material that the authors may not have intended to share.
However, beyond enabling quantification of sensitive information in source files, we find that accessing these URLs benefits authors by revealing resources that author potentially made publicly-accessible unintentionally.

We employed the following measures to further minimize harm:
\begin{enumerate*}[label=(\arabic*){}]
	\item For automated access (checking response codes), we adhered to established standards from Internet-wide scanning, including strict rate-limiting, a contact email as user-agent, an accompanying website served on the same host explaining the intent, and opt-out~\cite{Durumericetal2013ZMap:}.
	\item All access was read-only and strictly limited to the time needed to infer public read or write permissions and what type of content was accessible.
	\item As safeguards, we neither persisted any content nor transitively followed URLs, and we limited our manual assessment to the absolute minimum (\ie{} cloud storage services and collaborative platforms).
	\item We did not attempt to use or test any API tokens, private keys, or credentials found in online documents or any other source file.
	\item We responsibly disclosed all identified cases of potentially sensitive information to the affected authors to enable them to take corrective measures (see below).
	\item We do not publish any data or tooling that allows third parties to easily infer which submissions contain sensitive information.
\end{enumerate*}

\textbf{Responsible Disclosure.}
To minimize potential harm, we have conducted two responsible disclosure processes to allow affected stakeholders to timely react to our findings:

\textbf{\nostep{1}{} Inconsistency in Source Files Availability.}
\arxiv{} appeared to rely on two approaches for ``hiding'' source files (hiding the link vs.\ actually returning the HTTP Code 403), with only the latter resulting in the intended behavior.
We reached out to \arxiv{} on 2025-10-01 to inform them about this discrepancy.
\arxiv{} promptly informed us that this observation resulted from a bug, they had just fixed; \ie{} hiding these source file links was not intentional conduct.

\textbf{\nostep{2}{} Coordinated Information Campaign.}
Second, as summarized in \cref{subsubsec:analysis:individual:regexable}, we discovered hundreds of links to cloud services, websites, and file servers, some of which provided write access, and other sensitive information, such as private keys and tokens.
As detailed in the following, we first attempted to coordinate a disclosure process through \arxiv{}, but ultimately had to collect contact information (email addresses) and reach out to affected authors ourselves.

\textbf{Attempted Disclosure.}
For validated security-critical findings resulting from this data, we attempted to coordinate a disclosure process with \arxiv{}.
For affected submissions, we requested the submitters' email addresses.
However, this effort was unsuccessful because \arxiv{} is dedicated to protecting authors' identities even from ``trusted researchers such as [us]''.
Our subsequent attempts to let \arxiv{} disclose our findings to the authors on our behalf remained unanswered.
We attribute this behavior to \arxiv{} assigning responsibility ``primarily'' to the authors---disclosed to us via email and rather implicitly listed on the website~\cite{arXiv2005Requiring,arXiv2025Legacy}.
This view further emphasizes the importance of our research.

\textbf{Conducted Disclosure.}
We then applied a two-fold approach to disclose our findings.
First, we attempted to extract the email addresses of authors using regular expressions automatically.
Second, for papers where this attempt was unsuccessful, we manually retrieved them (retrieving the submitter's email address through \arxiv{} for \emph{all} impacted papers is not viable, as these requests are heavily rate-limited to combat spam).
Between 2025-11-04 and 2025-11-11, we reached out to, in total, \num{2660} authors of \num{1141} affected papers to
\begin{enumerate*}[label=(\roman*)]
    \item inform them about our findings, so that they could take corrective measures such as updating access permissions, and
    \item increase the awareness of corresponding issues and oversights.
\end{enumerate*}
Furthermore, we attached a request for participation in a short survey (see our research artifacts~\cite{Pennekampetal2026Artifact:}), which our institution's data protection officer approved.
As part of this survey, we did not collect any personal information, but participants could voluntarily provide their email addresses to follow up on this publication.

\textbf{Sensitivity Considerations.}
Given the potentially sensitive nature of the content we have been working with, we utilize a locally-hosted LLM (see \cref{app:sec:llmusage} for our sustainability-focused LLM usage considerations).
This way, we avoid revealing any sensitive data to third parties.

\textbf{Open Science Considerations.}
To balance our commitment to open science and the need to protect individual authors when releasing our analysis results~\cite{Pennekampetal2026Artifact:}, we obfuscate the paper identifiers while preserving the general identifier structure.
To prevent potential misuse of our artifacts, we explicitly decided against publishing our analysis code.

By carefully calibrating the information we release, responsibly disclosing identified security-critical findings before submission, and providing a new mitigation approach that addresses crucial shortcomings of existing tools, we maximize the benefits of our research for \arxiv{} authors/submitters and \LaTeX{} users while minimizing potential harm.

\newpage

\section{LLM Usage Considerations}
\label{app:sec:llmusage}

We locally-hosted an open-weights LLM to analyze comment contents at scale (\cf{} \cref{subsubsec:analysis:individual:comments}), preventing the disclosure of sensitive data to third parties.
To maintain accuracy, we carefully validated the LLM's outputs against human annotations (see \cref{subsubsec:analysis:individual:comments}) and emphasize that using an open-weights LLM has no implications on reproducibility.
Experiments ran on private infrastructure powered by solar energy during daytime operations (carbon efficiency: 0.25~kgCO$_2$eq/kWh).
We performed \qty{580}{\hour} of computation on two NVIDIA RTX B6000 GPUs (300W TDP), resulting in estimated total emissions of 87~kgCO$_2$eq~\cite{Lacosteetal2019Quantifying}.
The LLM was not used for content generation or literature review.

\section{Security Conference Submissions}
\label{app:sec:security}

To evaluate whether submissions connected to security conferences show different behavior regarding ``hidden'' information (\cf{} \cref{subsubsec:analysis:bigpicture:differences}), we identify matches between \arxiv{} submissions and papers published at security conferences by comparing paper titles based on information provided by dblp~\cite{Schloss-Dagstuhl1993dblp:}.
Utilizing the field of research code 4604 for cybersecurity and privacy, we selected all A\textsuperscript{*}, A, B, and C-ranked conferences from the CORE2023 ranking~\cite{Computing-Research--Education2013ICORE}.
We then applied a simple, case-insensitive matching (discarding all non-alphabetic characters) between paper titles on \arxiv{} and provided by dblp~\cite{Schloss-Dagstuhl1993dblp:} for these conferences.
Overall, we can only match a small set of \num{3289} security conference papers to \arxiv{} submissions.
We expected this range, especially since security conferences gradually shifted to quicker publication models (\eg{} hosting accepted versions papers immediately after acceptance or making the final versions available before the conference takes place) and paper titles can differ.
Specifically looking at A\textsuperscript{*} (\num{6}) and A (\num{11}) conferences, USENIX Sec. (\num{555} papers) and CCS (\num{472}) are most represented and ASIACRYPT (\num{54}) and CHES (\num{19}) least represented.
Relatively, PETS (\qty{26}{\percent} of papers) and EuroS\&P (\qty{21}{\percent}) have the highest and ASIACRYPT (\qty{3.1}{\percent}) and CHES (\qty{1.7}{\percent}) the lowest shares.

\section{Supplemental Evaluation Material}
\label{app:sec:evalextra}

We now provides further insights into the collected data.

\begin{table}
	\centering
	\caption{%
		Overview of regex-based pattern matches across \arxiv{} submissions highlights the extent of ``hidden'' secrets.
	}
	\vspace{-1.0em}
	\begin{scriptsize}
	\setlength{\tabcolsep}{3pt}
	\begin{tabularx}{\linewidth}{@{}
		X
		r
		r
		r
		r
		@{}
	}
		\toprule
		\textbf{Pattern} & \makecell[r]{\textbf{Estimated} \\ \textbf{Impact}} & \textbf{Submissions} & \makecell[r]{\textbf{Total Matches} \\ \textbf{(\% dang.)}} & \makecell[r]{\textbf{Valid} \\ \tiny\textbf{(Confirmed)}} \\
		\midrule

		Email addresses & \impM & 2.1\,M & 10.5\,M (\qty{30}{\percent}) & 5.8\,M--7.0\,M \\
		URLs & \impL & 1.7\,M & 680.3\,M (\qty{20}{\percent}) & 72.9\,M--123.5\,M \\
		Profanity & \impL & 1.1\,M & 6.5\,M (\qty{100}{\percent}) & 1.4\,M--1.7\,M \\
		P.O. box & \impL & 171\,K & 298\,K (\qty{33}{\percent}) & 268\,K--283\,K \\
		IPv4 addresses & \impL & 101\,K & 774\,K (\qty{17}{\percent}) & 46\,K--84\,K \\
		Credit card number & \impH & 17\,K & 1.9\,M (\qty{53}{\percent}) & 2--27\,K \\
		US SSN & \impH & 3.8\,K & 6.7\,K (\qty{13}{\percent}) & 79--277 \\
		Review form & \impM & 3.2\,K & 5.6\,K (\qty{52}{\percent}) & 2.5\,K--3.0\,K \\
		Generic passwords & \impH & 735 & 2.5\,K (\qty{52}{\percent}) & 456--632 \\
		AI Disclaimers & \impL & 537 & 1.2\,K (\qty{24}{\percent}) & 389--516 \\
		IBAN & \impM & 445 & 828 (\qty{24}{\percent}) & 21--45 \\
		censor pkg usage & \impM & 269 & 2.1\,K (\qty{23}{\percent}) & 1.8\,K--1.9\,K \\
		AWS access keys & \impH & 202 & 477 (\qty{27}{\percent}) & 92--123 \\
		Prompt injection & \impH & 162 & 498 (\qty{25}{\percent}) & 359--434 \\
		Bitcoin addresses & \impL & 103 & 1.3\,K (\qty{7}{\percent}) & 741--909 \\
		Password in URL & \impH & 82 & 171 (\qty{25}{\percent}) & 52--73 \\
		Bitcoin Bech32 & \impL & 25 & 115 (\qty{14}{\percent}) & 86--97 \\
		Google API keys & \impH & 24 & 99 (\qty{87}{\percent}) & 78 \\
		GitLab tokens & \impH & 18 & 42 (\qty{29}{\percent}) & 4 \\
		Slack tokens & \impH & 12 & 91 (\qty{45}{\percent}) & 3 \\
		Generic API keys & \impH & 10 & 18 (\qty{83}{\percent}) & 18 \\
		GitHub tokens & \impH & 10 & 21 (\qty{86}{\percent}) & 19 \\
		JWT tokens & \impM & 8 & 25 (\qty{16}{\percent}) & 24 \\
		/etc/passwd entries & \impM & 6 & 11 (\qty{27}{\percent}) & 9 \\
		Nmap scans & \impL & 6 & 16 (\qty{38}{\percent}) & 14 \\
		Hugging Face keys & \impH & 5 & 8 (\qty{62}{\percent}) & 5 \\
		Facebook OAuth & \impH & 5 & 5 (\qty{100}{\percent}) & 0 \\
		SSH private keys & \impH & 4 & 4 (\qty{50}{\percent}) & 4 \\
		OpenAI API keys & \impH & 3 & 18 (\qty{100}{\percent}) & 18 \\
		X access tokens & \impH & 3 & 4 (\qty{100}{\percent}) & 0 \\
		Google service acc. & \impH & 2 & 3 (\qty{67}{\percent}) & 3 \\
		Slack webhooks & \impH & 2 & 3 (\qty{33}{\percent}) & 2 \\
		/etc/shadow & \impM & 2 & 3 (\qty{0}{\percent}) & 3 \\
		Generic secrets & \impH & 1 & 1 (\qty{0}{\percent}) & 1 \\
		Google access tokens & \impH & 1 & 1 (\qty{0}{\percent}) & 1 \\
		\bottomrule
	\end{tabularx}
	\end{scriptsize}
	\vspace{-1.25em}
	\label{tab:appendix:regex-matches}
\end{table}

\textbf{Pattern Matching.}
\Cref{tab:appendix:regex-matches} gives an overview of our regex-based pattern-matching within \arxiv{} source files, which relies on prior collections (profanity~\cite{Ahn2009Offensive/Profane} and secrets pattern database~\cite{Ahmed2023Secrets}).
To check accuracy of these patterns we manually annotated up to 100 randomly sampled matches per pattern (``valid'' column) and estimate \qty{95}{\percent} confidence intervals for more frequent patterns via bootstrapping (n=\num{10000}).
As in \cref{subsec:analysis:individual}, we label patterns as potentially high (\impH), medium (\impM), and low (\impL) impact.
Our findings show that the majority of discoveries are not within dangling files, indicating that solely removing dangling files is insufficient to mitigate the disclosure of ``hidden'' information.
Instead, all source files have to be cleaned carefully.

\begin{figure}[t]
	\centering
	\includegraphics[width=\linewidth,trim={0 0 0 0},clip]{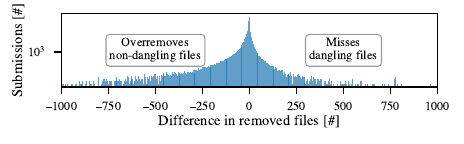}
	\vspace{-2.5em}
	\caption{%
		Distribution of over- and underremoved dangling files when applying cleaners that can remove files against our dangling file detection.
	}
	\vspace{-1.25em}
	\label{fig:cleaner-performance}
\end{figure}

\textbf{ALC Performance.}
\Cref{fig:cleaner-performance} further highlights the implications of relying on regex-based detection of dangling files as used in arxiv\_latex\_cleaner (ALC)~\cite{Google-Research2019arxivlatexcleaner}.
In addition to \emph{underremoving} files, and thereby keeping potentially sensitive files within ``cleaned'' sources, the \emph{overremoval} of files is a significant issue for usability, since the paper cannot be compiled anymore, possibly leading to authors ``simply'' providing the unsanitized source files instead.
Certainly, for many papers, the regex-based detection yields proper results; however, \qty{50}{\percent} papers of the \qty{2.32}{\million} evaluated submissions suffer from unintended results, featuring either over- or underremoval of dangling files.
The situation is further exacerbated by ALC's similarly suboptimal performance in our conducted comment cleanup tests (\cf{} \cref{tab:source-cleaner}).

\begin{figure}[t]
	\centering
	\includegraphics[width=\linewidth,trim={0 0 0 0.75em},clip]{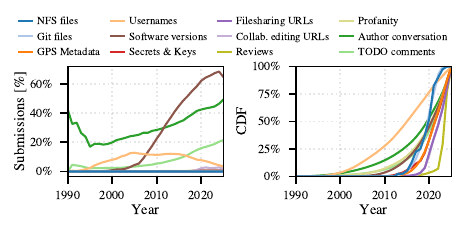}
	\vspace{-2.75em}
	\caption{%
		Longitudinal analysis of the categories of ``hidden'' information analyzed in \cref{subsec:analysis:individual} normalized by submissions per year (left), and frequency of individual categories (right) shows increasing occurrences.
	}
	\vspace{-1.25em}
	\label{fig:longitudinal}
\end{figure}

\textbf{Recent Submissions and Longitudinal Trends.}
We further complement the information shown in \cref{fig:issues-overview} (\cref{subsubsec:analysis:bigpicture:differences}) by adding a zoomed-in version (separated by months) of \arxiv{} submissions for 2025.
We do not see any notable trends across the different months, indicating largely consistent author behavior over time.
Moreover, as outlined in \cref{subsec:discussion:arxiv}, \arxiv{}'s new submission system (1.5), which queries for root-level dangling files, had no relevant impact on the broader picture of ``hidden'' information in source files so far.
Complementing our focus on 2025, \Cref{fig:longitudinal} visualizes the categories of ``hidden'' information detailed in \cref{subsec:analysis:individual} over time (1990--2025), showing a recent trend toward more and broader inclusion of such information.

\textbf{``Hidden'' Information in Our Own Papers.}
Out of curiosity, we also searched for unintentional disclosure of ``hidden'' information in our own \num{51} submissions at \arxiv{}.
\num{19} of them feature dangling files, which is reduced to \num{11} when only considering unique dangling files.
While we applied sanitization tools (mostly ALC) to some sources, we never sanitized any metadata before submitting our sources.
Overall, \num{13241} comments persist across our submissions, of which the LLM attributes \qty{6.4}{\percent} as private information, \qty{28}{\percent} as author conversation, and \qty{1.2}{\percent} as todos.
When applying \alcpp{} on our submissions, it removes \num{142} files and modifies \num{76} files when removing irrelevant content.
Fortunately, we did not discover any major issues among our own submissions.

\textbf{Anecdote: Sanitizing the \mbox{GPT-4} Paper.}
Taking another look at the paper on \mbox{GPT-4} that sparked discussions on Twitter/X~\cite{DV25591069650762023Sparks} (\cf{} \cref{sec:introduction}), we discovered that ALC requires manual intervention when cleaning the respective sources.
Out of the \num{206} dangling files (\qty{77}{\percent}), \num{46} \LaTeX{} files and \num{76} image/PDF files, it overremoves a file that is needed to compile the PDF.
Moreover, even though ALC reliably removes all dangling files and comments on intellectual property and toxicity that initially raised the discussions~\cite{DV25591069650762023Sparks}, it still underremoves other dangling files, including PDF files, and even overremoves content in \texttt{listings} environments, thereby effectively altering the paper's content.
In contrast, \alcpp{} accurately removes contained dangling files while outputting compilable, pixel-perfect \LaTeX{} source files.

\newpage

\section{Meta-Review}

The following meta-review was prepared by the program committee for the 2026 IEEE Symposium on Security and Privacy (S\&P) as part of the review process as detailed in the call for papers.

\subsection{Summary}
\label{app:metareview:summary}

This paper presents a large-scale empirical study of hidden information contained in the source files of \arXiv{} submissions.
By analyzing \qty{2.7}{\million} million submissions, the authors systematically identify unused files, embedded metadata, comments, and other artifacts that do not affect the compiled PDF but may contain sensitive information.
The paper further evaluates existing LaTeX sanitization tools and introduces a new open-source tool, \alcpp{}~\cite{Pennekampetal2026ALC-NG}, designed to more effectively remove hidden content.

\subsection{Scientific Contributions}
\begin{itemize}
\item Addresses a long-known issue.
\item Creates a new tool to enable future science.
\item Provides a valuable step forward in an established field.
\end{itemize}

\newpage
\vspace*{3.25em}

\subsection{Reasons for Acceptance}
\begin{enumerate}
\item While hidden information in \arXiv{} source files has been discussed anecdotally, this paper provides the first large-scale systematic study of the problem.
The authors quantify its prevalence across the corpus and identify concrete sensitive disclosures, including editable document links, API tokens, Git-embedded secrets, and GPS metadata.
\item The paper introduces \alcpp{}, a sanitization tool that improves on existing LaTeX cleaners by incorporating metadata analysis and structured source parsing.
The evaluation demonstrates clear limitations of prior tools and measurable improvements.
\item Beyond large-scale measurement, the authors conducted responsible disclosure and a follow-up survey.
The work advances understanding of privacy risks in academic publishing workflows and highlights gaps in current infrastructure practices.
\end{enumerate}

\end{document}